\documentclass{iopart}
\usepackage{amssymb}
\usepackage{iopams}
\usepackage{amsthm}
\usepackage{amscd}
\usepackage{amsfonts}
\usepackage{amsbsy}
\catcode`\@=11
\renewcommand\footnoterule{%
  \kern-3\p@
  \hrule\@width.4\columnwidth
  \kern2.6\p@}
\renewcommand\@makefntext[1]{%
    \parindent 1em\noindent
    \hb@xt@1.8em{\hss$^{\@thefnmark}$)}\hspace{2pt}%
    \footnotesize\rmfamily#1}  
\def\@makefnmark{\hspace{.5pt}\hbox{$^{\@thefnmark}$%
\hspace{-1pt})}} 
\catcode`\@=11
\renewcommand\footnoterule{%
  \kern-3\p@
  \hrule\@width.4\columnwidth
  \kern2.6\p@}
\renewcommand\@makefntext[1]{%
    \parindent 1em\noindent
    \hb@xt@1.8em{\hss$^{\@thefnmark}$)}\hspace{2pt}%
    \footnotesize\rmfamily#1}  
\def\@makefnmark{\hspace{.5pt}\hbox{$^{\@thefnmark}$%
\hspace{-1pt}}} 

 \newtheorem{ttt}{\bfseries{Theorem}}[section]
 \newtheorem{ddd}{\bfseries{Definition}}[section]
 \newtheorem{ppp}{\bfseries{Proposition}}[section]
 \newtheorem{lelele}{\bfseries{Lemma}}[section]
 \newtheorem{ccc}{\bfseries{Corollary}}[section]

\def\RR{\mathbb{R}}
\def\NN{\mathbb{N}}

\def\a{\alpha}
\def\b{\beta}

\def\sg{\sigma}

\def\om{\omega}

\newcommand{\cC}{\mathcal{C}}

\newcommand{\cP}{\mathcal{P}}
\newcommand{\cT}{\mathcal{T}}
\newcommand{\cQ}{\mathcal{Q}}

\newcommand{\be}[1]{\begin{equation}\label{#1}}
\newcommand{\ee}{\end{equation}}
\newcommand{\ba}[1]{\begin{eqnarray}\label{#1}}
\newcommand{\ea}{\end{eqnarray}}
\newcommand{\rf}[1]{(\ref{#1})}
\newcommand{\nn}{\nonumber}

\newcommand{\sign}{\mbox{\rm sign}\,}

\begin{document}
\title[$J$-self-adjoint operators with $\mathcal{C}$-symmetries]{$J$-self-adjoint operators with $\mathcal{C}$-symmetries: extension theory approach}

\author{S. Albeverio${}^a$,\ U. G\"{u}nther${}^b$,\  and S. Kuzhel${}^c$}

\address{${}^a$\ Institut f\"{u}r Angewandte Mathematik,
Universit\"{a}t Bonn, Wegelerstr. 6, D-53115 Bonn, Germany; SFB 611
and HCM, Bonn, BiBoS, Bielefeld-Bonn, Germany; CERFIM, Locarno and
USI
Switzerland\\
${}^b$\ Forschungszentrum Dresden-Rossendorf, PO 510119, D-01314
Dresden, Germany\\
${}^c$\ Institute of Mathematics of the National Academy of Sciences
of Ukraine, Tereshchenkovskaya 3, 01601 Kyiv, Ukraine}
\eads{\mailto{albeverio@uni-bonn.de},\ \mailto{u.guenther@fzd.de},\
\mailto{kuzhel@imath.kiev.ua}}
\begin{abstract}
A well known tool in conventional (von Neumann) quantum mechanics is
the self-adjoint extension technique for symmetric operators. It is
used, e.g., for the construction of Dirac-Hermitian Hamiltonians
with point-interaction potentials. Here we reshape this technique to
allow for the construction of pseudo-Hermitian ($J$-self-adjoint)
Hamiltonians with complex point-interactions. We demonstrate that
the resulting Hamiltonians are bijectively related with so called
hypermaximal neutral subspaces of the defect Krein space of the
symmetric operator. This symmetric operator is allowed to have
arbitrary but equal deficiency indices $<n,n>$. General properties
of the $\cC$ operators for these Hamiltonians are derived. A
detailed study of  $\cC$-operator parametrizations  and Krein type
resolvent formulas is provided for $J$-self-adjoint extensions of
symmetric operators with deficiency indices $<2,2>$. The technique
is exemplified on 1D pseudo-Hermitian Schr\"odinger and Dirac
Hamiltonians with complex point-interaction potentials.
\end{abstract}





\section{Introduction}
The use of non-Hermitian operators and indefinite Hilbert space
structures in quantum mechanics dates back to the early 1940s
\cite{Di,Pa}. The interest in this subject strongly increased after
it has been discovered in 1998 that complex Hamiltonians possessing
$\mathcal{PT}$-symmetry (the product of parity and time reversal)
can have a real spectrum (like self-adjoint operators) \cite{B1}.
This gave rise to a consistent complex extension of conventional
quantum mechanics (CQM) into $\mathcal{PT}$ quantum mechanics
(PTQM), see e.g. the review paper \cite{B4} and the references
therein.

During the past ten years PTQM models have been analyzed with a
wealth of technical tools (for an overview see
\cite{PT-SI2,PT-SI3,PT-SI4,PT-SI5}). Most prominent ones concern
Bethe Ansatz techniques (to prove the reality of the spectrum for
the Hamiltonian with complex cubic potential $ix^3$ which originated
a lot of interest) \cite{DDT}, various global approaches based on
the extension of differential operators into the complex coordinate
plane \cite{BBM-jmp,BCDM,smilga,MZ-tobo}, SUSY approaches
\cite{susy-1,susy-2,susy-3,susy-4}, $\cP\cT-$symmetric perturbations
of Hermitian operators \cite{cal-pert}, Moyal-product
\cite{moyal-1,moyal-2} and Lie-algebraic \cite{lie-algebra}
techniques. We would also like to mention the more recent
considerations on spectral degeneracies
\cite{EP-GRS,EP-MZ,EP-cali,EP-GGKN}.

Apart from these techniques and applications, one of the most
important concepts to place ${\mathcal P}{\mathcal T}$-symmetry in a
general mathematical context remains the concept of
\emph{pseudo-Hermiticity} \cite{M1}. A linear densely defined
operator $A$ acting in a Hilbert space ${\mathfrak H}$ with the
inner product $(\cdot, \cdot)$ is called pseudo-Hermitian if its
adjoint ${A}^*$ satisfies the condition
\begin{equation}\label{e1}
{A}^*\eta=\eta{A},
\end{equation}
where $\eta$ is an invertible bounded self-adjoint operator in
${\mathfrak H}$. Since a Hilbert space $\mathfrak{H}$ endowed with
an indefinite metric $[f,g]_{\eta}=(\eta{f},g)$ is an example of a
Krein space with fundamental symmetry $J=\eta|\eta|^{-1}$ (here
$|\eta|=\sqrt{\eta^2}$ is the modulus of $\eta$) \cite{AZ,DL}, one
can reduce the investigation of pseudo-Hermitian operators to the
study of $J$-self-adjoint operators in a Krein space
\cite{Ja,AK2,LT,GSZ,TT}.

We recall that a linear densely defined operator $A$ acting in a
Krein space $({\mathfrak H}, [\cdot, \cdot]_J)$ with fundamental
symmetry $J$ (i.e., $J=J^*$ and $J^2=I$) and indefinite metric
$[\cdot, \cdot]_J=(J\cdot,\cdot)$ is called $J$-{\it self-adjoint}
if ${A}^*J=J{A}$. Obviously, $J$-self-adjoint operators are
pseudo-Hermitian ones in the sense of (\ref{e1}).

In contrast to self-adjoint operators in Hilbert spaces (which
necessarily have a purely real spectrum), self-adjoint operators in
Krein spaces, in general, have a spectrum which is only symmetric
with respect to the real axis \cite{AZ,DL}. Pairwise complex
conjugate eigenvalues, as part of the discrete spectrum,  are
connected with spontaneously broken $\cP\cT-$symmetry. This means
that although the Hamiltonian will have $\cP\cT-$symmetry its
eigenfunctions will not be $\cP\cT-$symmetric. The real discrete
spectrum corresponds to the sector of so-called exact
$\cP\cT-$symmetry where in addition to the Hamiltonian also its
eigenfunctions are $\cP\cT-$symmetric.

One of the key points in PTQM is the description of a hidden
symmetry $\mathcal{C}$ \cite{B-C} which is present for a given
${\mathcal P}{\mathcal T}$-symmetric Hamiltonian $A$ in the sector
of exact $\cP\cT-$symmetry.

By analogy with \cite{B4} the definition of $\mathcal{C}$-symmetry
for the case of $J$-self-adjoint operators can be formalized as
follows.
\begin{ddd}\label{dad1}
A $J$-self-adjoint operator ${A}$ has the property of
$\mathcal{C}$-symmetry if there exists a bounded linear operator
$\mathcal{C}$ in $\mathfrak{H}$ such that: \ $(i) \
{\mathcal{C}}^2=I;$ \quad $(ii) \ J\mathcal{C}>0$; \quad $(iii) \
A{\mathcal{C}}={\mathcal{C}}A$.
\end{ddd}

The properties of $\mathcal{C}$ are nearly identical to those of the
charge conjugation operator in quantum field theory and the
existence of $\mathcal{C}$ provides an inner product
$(\cdot,\cdot)_{{\mathcal{C}}}=[{\mathcal{C}}\cdot,\cdot]_J$ whose
associated norm is positive definite and the dynamics generated by
$A$ is therefore governed by a unitary time evolution. However, the
operator $\mathcal{C}$ depends on the choice of $A$ and its finding
is a nontrivial problem \cite{B,BJ,BT,M-metric}. A generalization
from bounded to unbounded $\cC$ operators was recently discussed in
\cite{K-C}. Another kind of generalized $\cC$ operator can arise in
connection with model classes of interacting relativistic quantum
fields with indefinite metrics and satisfying all Morchio-Strocchi
axioms, see, e.g. \cite{AG-cmp-2001} (and references therein).

In the present paper, we are going to study $J$-self-adjoint
operators with $\mathcal{C}$-symmetries within an extension theory
approach. This means that the complex potential, typical for PTQM
Hamiltonians, is induced via point-interactions which are described
by an operator extension technique. The extension technique is a
standard mathematical tool \cite{RS} in CQM and is widely used to
efficiently describe point interactions \cite{AL,AL1}. PTQM-related
considerations based on this technique can be found in
\cite{AK2,A3,AK}.

The paper is structured as follows. Section 2 contains an abstract
study of $\mathcal{C}$-symmetries in a Krein space approach and has
an auxiliary character. In Section 3, we describe all
$J$-self-adjoint extensions of a given symmetric operator
$A_{\mathrm{sym}}$ (under the condition
$A_{\mathrm{sym}}J=JA_{\mathrm{sym}}$) and, for the case of
deficiency indices $<2,2>$, we propose a general method allowing us:
(i) to describe the set of $J$-self-adjoint extensions $A_{M(U)}$ of
$A_{\mathrm{sym}}$ with $\mathcal{C}$-symmetries; (ii) to construct
the corresponding $\mathcal{C}$-symmetries in a simple explicit form
(family of $C_{\theta,\omega}$-symmetries); (iii) to establish a
Krein-type resolvent formula for $J$-self-adjoint extensions
$A_{M(U)}$ with $\mathcal{C}$-symmetries. Section 4 illustrates the
obtained results on the examples of a Schr\"{o}dinger operator with
general zero-range potential and a one-dimensional Dirac Hamiltonian
with point perturbation.

Let us briefly comment on the used notations. $\mathcal{D}(A)$ and
$\mathcal{R}(A)$ denote the domain and the range of a linear
operator $A$, respectively. $A\upharpoonright{\mathcal{D}}$ means
the restriction of $A$ onto a set $\mathcal{D}$.

\setcounter{equation}{0}

\section{$J$-Self-Adjoint Operators with  $\mathcal{C}$-Symmetries}
\subsection{Elements of Krein space theory.}

Here all necessary results of Krein space theory are presented in a
form convenient for our exposition. Their proofs and detailed
analysis can be found in \cite{AZ,DL}.

Let ${\mathfrak H}$ be a Hilbert space with inner product
$(\cdot,\cdot)$ and with fundamental symmetry (involution) $J$
(i.e., $J=J^*$ and $J^2=I$). The corresponding orthoprojectors
$P_+=1/2(I+J)$, $P_-=1/2(I-J)$ determine the fundamental
decomposition of ${\mathfrak H}$
\begin{equation}\label{d1}
{\mathfrak H}={\mathfrak H}_+\oplus{\mathfrak H}_-, \qquad
{\mathfrak H}_-=P_{-}{\mathfrak H}, \quad {\mathfrak
H}_+=P_{+}{\mathfrak H}.
\end{equation}

The space ${\mathfrak H}$ endowed with the indefinite inner product
(indefinite metric)
\begin{equation}\label{e29}
 [x,y]_J:=(J{x}, y), \qquad \forall{x,y}\in\mathfrak{H}
\end{equation}
is called {\it a Krein space}  $({\mathfrak H}, [\cdot,\cdot]_J)$.

A subspace $\mathfrak{L}\subset{\mathfrak H}$ is called hypermaximal
neutral if $\mathfrak{L}$ coincides with its $J$-orthogonal
complement: $
\mathfrak{L}=\mathfrak{L}^{[\bot]_J}=\{x\in\mathfrak{H}
 :  [x,y]_J=0, \ \forall{y}\in\mathfrak{L}\}$. Hypermaximal neutral
subspaces exist only in the case where $\dim{\mathfrak
H}_+=\dim{\mathfrak H}_-$.

A subspace $\mathfrak{L}\subset{\mathfrak H}$ is called {\it
nonnegative, positive, uniformly positive} if, respectively,
$[x,x]_J\geq0, \quad [x,x]_J>0, \quad [x,x]_J\geq\alpha^2\|x\|^2$,
$\a\in\RR$ for all $x\in\mathfrak{L}\setminus\{0\}$. Nonpositive,
negative, and uniformly negative subspaces are introduced
similarly. The subspaces $\mathfrak{H}_{\pm}$ in (\ref{d1}) are
examples of uniformly positive and uniformly negative subspaces
and they possess the property of maximality in the corresponding
classes (i.e., $\mathfrak{H}_{+}$ ($\mathfrak{H}_{-}$) does not
belong as a subspace to any uniformly positive (negative)
subspace).

Let a subspace $\mathfrak{L}_+$ be maximal uniformly positive. Then
its $J$-orthogonal complement $
\mathfrak{L}_-=\mathfrak{L}_+^{[\bot]_J}$ is a maximal uniformly
negative subspace of $\mathfrak{H}$ and the direct $J$-orthogonal
sum
\begin{equation}\label{d2}
{\mathfrak H}={\mathfrak L}_+[\dot{+}]_J{\mathfrak L}_-
\end{equation}
gives the decomposition of ${\mathfrak H}$ onto its positive
${\mathfrak L}_+$ and negative ${\mathfrak L}_-$ parts (the brackets
$[\cdot]_J$ mean the orthogonality with respect to the indefinite
metric).

The subspaces ${\mathfrak L}_+$ and ${\mathfrak L}_-$ in (\ref{d2})
can be described as ${\mathfrak L}_+=(I+K)\mathfrak{H}_+$ and
${\mathfrak L}_-=(I+Q)\mathfrak{H}_-$, where
$K:\mathfrak{H}_+\to\mathfrak{H}_-$ is a contraction and
$Q=K^*:\mathfrak{H}_-\to\mathfrak{H}_+$ is the adjoint of $K$.

The self-adjoint operator $T=KP_++K^*P_-$ acting in $\mathfrak{H}$
is called {\it an operator of transition} from the fundamental
decomposition (\ref{d1}) to (\ref{d2}). Obviously,
\begin{equation}\label{d4}
{\mathfrak L}_+=(I+T)\mathfrak{H}_+, \qquad {\mathfrak
L}_-=(I+T)\mathfrak{H}_-.
\end{equation}
Furthermore, the projectors $P_{\mathfrak{L}_\pm} :
\mathfrak{H}\to{\mathcal L}_{\pm}$ onto $\mathfrak{L}_\pm$ with
respect to the decomposition (\ref{d2}) have the form
\begin{equation}\label{k1}
P_{\mathfrak{L}_-}=(I-T)^{-1}(P_--TP_+),  \quad
P_{\mathfrak{L}_+}=(I-T)^{-1}(P_+-TP_-).
\end{equation}

The collection of operators of transition admits a simple `external'
description. Namely, a self-adjoint operator $T$ in  $\mathfrak{H}$
is an operator of transition if and only if
\begin{equation}\label{e3}
\|T\|<1  \qquad \mbox{and} \qquad JT=-TJ.
\end{equation}

\subsection{$J$-Self-adjoint operators with  $\mathcal{C}$-symmetries.}

The next statement characterizes the structure of $J$-self-adjoint
operators with $\mathcal{C}$-symmetries.

\begin{ttt}[\cite{AK2}]\label{t1}
A $J$-self-adjoint operator $A$ acting in a Krein space $({\mathfrak
H}, [\cdot, \cdot]_J)$ has the property of ${\mathcal C}$-symmetry
if and only if $A$ admits the decomposition
\begin{equation}\label{new1}
A=A_+{[\dot{+}]_J}A_-, \qquad
A_{+}=A\upharpoonright\mathfrak{L}_{+}, \quad
A_{-}=A\upharpoonright\mathfrak{L}_{-}
\end{equation}
with respect to a certain choice of the $J$-orthogonal decomposition
(\ref{d2}) of $\mathfrak{H}$. In that case
\begin{equation}\label{e2}
{\mathcal C}=P_{\mathfrak{L}_+}-P_{\mathfrak{L}_-}=(I+T)(I-T)^{-1}J,
\end{equation}
where $T$ is the operator of transition from the fundamental
decomposition (\ref{d1}) to (\ref{d2}).
\end{ttt}

\noindent {\bf Remark 2.1}
 Since $T$ is a self-adjoint operator and $\|T\|<1$, the formula
(\ref{e2}) can be rewritten as ${\mathcal C}=e^{Q}J$, where $Q$
$(=\ln{(I+T)(I-T)^{-1}})$ is a bounded self-adjoint operator in
$\mathfrak{H}$. Then the condition ${\mathcal{C}}^2=I$ takes the
form $e^{Q}J=Je^{-Q}$ which implies $QJ=-JQ$. Therefore, one can
rewrite (\ref{e2}) as
\begin{equation}\label{aaa3}
\mathcal{C}=e^{Q}J=e^{Q/2}Je^{-Q/2}.
\end{equation}

Set $(\cdot,\cdot)_\mathcal{C}\equiv [\mathcal{C}\cdot,\cdot]_J$.
Due to (\ref{aaa3}),
$(\cdot,\cdot)_\mathcal{C}=(e^{-Q/2}\cdot,e^{-Q/2}\cdot)$. The
sesquilinear form $(\cdot,\cdot)_\mathcal{C}$ determines a new inner
product in $\mathfrak{H}$ that is equivalent to the initial one.
Since ${\mathcal C}=P_{\mathfrak{L}_+}-P_{\mathfrak{L}_-}$ (by
(\ref{e2})), the $J$-orthogonal decomposition (\ref{d2}) is
transformed into the orthogonal sum $\mathfrak{H}={\mathfrak
L}_+\oplus_{\mathcal C}{\mathfrak L}_-$ with respect to the inner
product $(\cdot,\cdot)_\mathcal{C}$ and the decomposition
(\ref{new1}) takes the form $A=A_+{\oplus_{\mathcal C}}A_-$.

\begin{ccc}\label{c1}
Let $A$ be a $J$-self-adjoint operator. The following statements are
equivalent:

(i) $A$ has the property of ${\mathcal C}$-symmetry;

(ii) the operators $A_{+}$ and $A_{-}$ in the decomposition
$A=A_+{\oplus_{\mathcal C}}A_-$ are self-adjoint in the Hilbert
spaces $\mathfrak{L}_{+}$ and $\mathfrak{L}_{-}$ with the inner
product  $(\cdot,\cdot)_{\mathcal{C}}$;

(iii) the operator $H={e^{-Q/2}}A{e^{Q/2}}$ is self-adjoint in
$\mathfrak{H}$;
\end{ccc}

{\it Proof.} By (\ref{e2}) the restriction of
$(\cdot,\cdot)_\mathcal{C}$ onto the subspaces $\mathfrak{L}_+$ and
$\mathfrak{L}_-$ coincides with $[\cdot, \cdot]_J$ and $-[\cdot,
\cdot]_J$, respectively. This means that the assumption of
$J$-self-adjointness of $A$ is equivalent to the property of
self-adjointness of $A_{\pm}=A\upharpoonright\mathfrak{L}_{\pm}$
with respect to $(\cdot,\cdot)_\mathcal{C}$. Hence, $(i)\iff(ii)$.

By virtue of $(ii)$,  $A$ is self-adjoint in $\mathfrak{H}$ with
respect to the inner product $(\cdot,\cdot)_\mathcal{C}$. Therefore,
$$
(e^{-Q/2}Ax,e^{-Q/2}y)=(Ax, y)_\mathcal{C}=(x,
Ay)_\mathcal{C}=(e^{-Q/2}x,e^{-Q/2}Ay), \quad
\forall{x,y}\in\mathfrak{H}.
$$
This means that the operator $H=e^{-Q/2}Ae^{Q/2}$ is self-adjoint in
$\mathfrak{H}$ with respect to the initial product $(\cdot,\cdot)$
if and only if $A$ is self-adjoint with respect to
$(\cdot,\cdot)_\mathcal{C}$. Thus $(ii)\iff(iii)$. \rule{2mm}{2mm}

\begin{ccc}\label{c23}
If a $J$-self-adjoint operator $A$ has the property of ${\mathcal
C}$-symmetry then its spectrum $\sigma(A)$ is real and the adjoint
operator $\mathcal{C}^*$ provides the property of
$\mathcal{C}$-symmetry for $A^*$.
\end{ccc}

{\it Proof.} The reality of $\sigma(A)$ follows from assertion
$(ii)$ of Corollary \ref{c1}. If $A$ has ${\mathcal C}$-symmetry,
then the adjoint $\mathcal{C}^*$ satisfies all conditions of
Definition \ref{dad1} for $A^*$. So, $\mathcal{C}^*$ provides the
property of $\mathcal{C}$-symmetry for $A^*$. \rule{2mm}{2mm}

\noindent {\bf Remark 2.2} In the context of PTQM, the existence of
an equivalence mapping (similarity transformation) $e^{Q/2}$ between
a pseudo-Hermitian operator $A$ and a Hermitian operator $H$ was
first demonstrated by Mostafazadeh in \cite{M2}. The $\cC$ operator
was introduced in PTQM by Bender, Brody and Jones in \cite{B-C}. As
it is obvious from \rf{aaa3}, $\cC$ as a dynamically adapted
($A-$dependent) involution is a similarity transformed  version of
the original involution $J$.

\setcounter{equation}{0}
\section{Extension Theory Approach}
\subsection{Preliminaries on extension theory. General case.}
Let $A_{\mathrm{sym}}$ be a closed symmetric densely defined
operator in $\mathfrak{H}$ with the equal deficiency indices $<n,n>$
\ $(n\in\mathbb{N}\cup\{\infty\})$. Denote by
$\mathfrak{N}_i=\mathfrak{H}\ominus\mathcal{R}(A_{\mathrm{sym}}+iI)$
and
$\mathfrak{N}_{-i}=\mathfrak{H}\ominus\mathcal{R}(A_{\mathrm{sym}}-iI)$
the defect subspaces of $A_{\mathrm{sym}}$ and consider the Hilbert
space $\mathfrak{M}=\mathfrak{N}_{-i}\dot{+}\mathfrak{N}_{i}$ with
the inner product
\begin{equation}\label{aaa2}\fl
(x,y)_{\mathfrak{M}}=2[(x_i,y_i)+(x_{-i},y_{-i})] \quad
x=x_i+x_{-i}, \quad y=y_i+y_{-i} \quad \{x_{{\pm}i},
y_{{\pm}i}\}\subset\mathfrak{N}_{{\pm}i}.
\end{equation}

Obviously, the operator $Z(x_i+x_{-i})=x_i-x_{-i}$ is a fundamental
symmetry in the Hilbert space $\mathfrak{M}$ and it acts as identity
operator on $\mathfrak{N}_{i}$ and minus identity operator on
$\mathfrak{N}_{-i}$.

In what follows we assume that
\begin{equation}\label{es12}
A_{\mathrm{sym}}J=JA_{\mathrm{sym}},
\end{equation}
where $J$ is a fundamental symmetry in $\mathfrak{H}$. Then the
subspaces $\mathfrak{N}_{\pm{i}}$ reduce $J$ and the restriction
$J\upharpoonright\mathfrak{M}$ gives rise to the fundamental
symmetry in the Hilbert space $\mathfrak{M}$. Moreover, according to
the properties of $Z$ mentioned above, $JZ=ZJ$ and $JZ$ is
 a fundamental symmetry in $\mathfrak{M}$. Therefore, the
sesquilinear form
\begin{equation}\label{e46}
[x,y]_{JZ}=(JZx,y)_{\mathfrak{M}}=2[(Jx_i,y_i)-(Jx_{-i},y_{-i})]
\end{equation}
defines an indefinite metric on $\mathfrak{M}$.

According to von-Neumann formulas any closed intermediate extension
$A$ of $A_{\mathrm{sym}}$ (i.e.,
${A}_{\mathrm{sym}}\subset{A}\subset{A}_{\mathrm{sym}}^*$) is
uniquely determined by the choice of a subspace
$M\subset\mathfrak{M}$. This means that
$\mathcal{D}(A)=\mathcal{D}(A_{\mathrm{sym}})\dot{+}M$ and
\begin{equation}\label{e55}\fl
Af=A_{\mathrm{sym}}^*(u+x)=A_{\mathrm{sym}}u+iZx, \quad
\forall{u}\in\mathcal{D}(A_{\mathrm{sym}}), \quad \forall{x}\in{M}.
\end{equation}

Taking (\ref{es12}) -- (\ref{e55}) into account we immediately
derive
\begin{equation}\label{e78a}\fl
[A_1f_1, f_2]_{J}-[f_1, A_2f_2]_{J}=i[x_1,x_2]_{JZ}, \quad
\forall{f_j=u_j+x_j}\in\mathcal{D}(A_{j}), \ x_j\in{M_j}
\end{equation}
for arbitrary intermediate extensions $A_1$ and $A_2$ of
$A_{\mathrm{sym}}$ which are defined by the subspaces $M_1$ and
$M_2$, respectively (see e.g., \cite[Lemma 9.6]{KK}).

It follows from (\ref{e78a}) that an extension
$A\supset{A_{\mathrm{sym}}}$ defined by $M$ is a $J$-self-adjoint
operator if and only if
$$
M={M}^{[\perp]_{JZ}}=\{y\in\mathfrak{M} \ : \ [x,y]_{JZ}=0, \
\forall{x}\in{M}\},
$$
i.e., if $M$ is a hypermaximal neutral subspace of the Krein space
$(\mathfrak{M}, [\cdot, \cdot]_{JZ})$.

The next statement is a `folklore' result of extension theory.

\begin{ppp}\label{t2}
Let $A_{\mathrm{sym}}J=JA_{\mathrm{sym}}$. Then the correspondence
$A\leftrightarrow{M}$ determined by (\ref{e55}) is a bijection
between the set of all $J$-self-adjoint extensions $A$ of
$A_{\mathrm{sym}}$ and the set of all hypermaximal neutral subspaces
$M$ of $(\mathfrak{M}, [\cdot, \cdot]_{JZ})$.
\end{ppp}

To underline the relationship $A\leftrightarrow{M}$ we will use the
notation $A_M$ for $J$-self-adjoint extensions $A$ of
$A_{\mathrm{sym}}$ determined by (\ref{e55}).
\begin{ttt}\label{p1}
Let $A_{\mathrm{sym}}J=JA_{\mathrm{sym}}$ and let
$A_{\mathrm{sym}}\mathcal{C}=\mathcal{C}A_{\mathrm{sym}}$, where
$\mathcal{C}$ is a bounded linear operator in $\mathfrak{H}$ such
that ${\mathcal{C}}^2=I$ and $J\mathcal{C}>0$. Then a
$J$-self-adjoint extension $A_M$ of $A_{\mathrm{sym}}$ has
$\mathcal{C}$-symmetry if and only if $\mathcal{C}M=M$.
\end{ttt}

{\it Proof.} Since $A_{\mathrm{sym}}$ commutes with $J$ and
$\mathcal{C}$ one gets
$A_{\mathrm{sym}}e^{Q}=e^{Q}A_{\mathrm{sym}}$, where the
self-adjoint operator $e^{Q}$ is defined in (\ref{aaa3}). But then $
A_{\mathrm{sym}}\mathcal{C}^{*}=A_{\mathrm{sym}}Je^{Q}=Je^{Q}A_{\mathrm{sym}}=\mathcal{C}^{*}A_{\mathrm{sym}}.$
The relations
$\mathcal{C}^{*}A_{\mathrm{sym}}=A_{\mathrm{sym}}\mathcal{C}^{*}$
and $\mathcal{C}^2=I$ imply
$\mathcal{C}\mathfrak{N}_{\pm}=\mathfrak{N}_{\pm}$ and hence,
$\mathcal{C}\mathfrak{M}=\mathfrak{M}$.

Using the identity
$\mathcal{C}A_{\mathrm{sym}}^{*}=A_{\mathrm{sym}}^{*}\mathcal{C}$
which immediately follows from
$\mathcal{C}^{*}A_{\mathrm{sym}}=A_{\mathrm{sym}}\mathcal{C}^{*}$
one concludes that
$\mathcal{C}A_M=A_M\mathcal{C}\iff\mathcal{C}\mathcal{D}(A_M)=\mathcal{D}(A_M)$.
Taking the relations
$\mathcal{D}(A_M)=\mathcal{D}(A_{\mathrm{sym}})\dot{+}M$,
$\mathcal{C}\mathcal{D}(A_{\mathrm{sym}})=\mathcal{D}(A_{\mathrm{sym}})$,
and $\mathcal{C}\mathfrak{M}=\mathfrak{M}$ into account one gets
$\mathcal{C}A_M=A_M\mathcal{C}\iff\mathcal{C}M=M$. Theorem \ref{p1}
is proved. \rule{2mm}{2mm}

\noindent {\bf Remark 3.1} The commutation relation
$A_{\mathrm{sym}}J=JA_{\mathrm{sym}}$ in theorem \ref{p1} is a
natural condition in the present approach because the
complex-potential properties of the $J-$self-adjoint operators $A$
are induced only by the boun\-da\-ry-con\-di\-tion-re\-la\-ted
extension families (see below).

\subsection{The case of deficiency indices $<2,2>$.}

In what follows we assume that the symmetric operator
$A_{\mathrm{sym}}$ has the deficiency indices $<2,2>$ and
\emph{there exists at least one $J$-self-adjoint extension $A_M$ of
$A_{\mathrm{sym}}$}. In that case $\dim\mathfrak{M}=4$ and each of
the orthogonal subspaces of $\mathfrak{M}$:
$$
\begin{array}{lr}
{\mathfrak{M}}_{++}=(I+Z)(I+J)\mathfrak{M}; \ & \ {\mathfrak{M}}_{--}=(I-Z)(I-J)\mathfrak{M}; \vspace{5mm} \\
 {\mathfrak{M}}_{+-}=(I+Z)(I-J)\mathfrak{M}; \ & \
{\mathfrak{M}}_{-+}=(I-Z)(I+J)\mathfrak{M}
\end{array}
$$
is one-dimensional. (Otherwise, $Z={J}$ or $Z=-J$ and there exist no
$J$-self-adjoint extensions of $A_{\mathrm{sym}}$ --- in
contradiction to the above assumption.)

Let $\{e_{\pm\pm}\}$ be an orthonormal basis of $\mathfrak{M}$ such
that ${\mathfrak{M}}_{\pm\pm}=<e_{\pm\pm}>$. It follows from the
definition of ${\mathfrak{M}}_{\pm\pm}$ that
\begin{equation}\label{ss1}\fl
\begin{array}{lr}
Je_{++}=e_{++}, \quad Je_{-+}=e_{-+}, & Je_{+-}=-e_{+-}, \quad Je_{--}=-e_{--};  \\
Ze_{++}=e_{++}, \quad Ze_{-+}=-e_{-+}, & Ze_{+-}=e_{+-}, \quad
Ze_{--}=-e_{--}.
\end{array}
\end{equation}

This means that the fundamental decomposition of the Krein space
$(\mathfrak{M}, [\cdot, \cdot]_{JZ})$ has the form
\begin{equation}\label{e5}\fl
\mathfrak{M}={\mathfrak{M}}_{-}\oplus{\mathfrak{M}}_{+}, \qquad
{\mathfrak{M}}_{-}=<e_{+-}, e_{-+}>, \quad
{\mathfrak{M}}_{+}=<e_{++}, e_{--}>.
\end{equation}

According to the general theory \cite{AZ}, an arbitrary hypermaximal
neutral subspace $M$ of $(\mathfrak{M}, [\cdot, \cdot]_{JZ})$ can be
uniquely determined by a unitary mapping of ${\mathfrak{M}}_{-}$
onto ${\mathfrak{M}}_{+}$. Since
$\dim{\mathfrak{M}}_{+}=\dim{\mathfrak{M}}_{-}=2$ the set of unitary
mappings ${\mathfrak{M}}_{-}\to{\mathfrak{M}}_{+}$ is determined by
the set of unitary matrices
\begin{equation}\label{eee6}\fl
U=e^{i\phi}\left(\begin{array}{cc} qe^{i\gamma} & re^{i\xi} \\
-re^{-i\xi} & qe^{-i\gamma} \end{array}\right), \qquad q^2+r^2=1,
\quad \phi, \gamma, \xi\in{[0,2\pi)}.
\end{equation}
(We have used the standard representation $U(2)=U(1)\times{SU(2)}$
for the reducible $U(2)$ group elements \cite{ENC}).

In other words, the decomposition (\ref{e5}) and representation
(\ref{eee6}) allow one to describe a hypermaximal neutral subspace
$M$ of $(\mathfrak{M}, [\cdot, \cdot]_{JZ})$ as a linear span
\begin{equation}\label{e6}
M=M(U)=<d_1,d_2>
\end{equation}
of elements
\begin{equation}\label{aaa5}
\begin{array}{c}
d_1=e_{++}+qe^{i(\phi+\gamma)}e_{+-}+re^{i(\phi+\xi)}e_{-+}; \\
d_2=e_{--}-re^{i(\phi-\xi)}e_{+-}+qe^{i(\phi-\gamma)}e_{-+}.
\end{array}
\end{equation}

By Proposition \ref{t2}, formula (\ref{e6}) provides a one-to-one
correspondence between domains
$\mathcal{D}(A_{M(U)})=\mathcal{D}(A_{\mathrm{sym}})\dot{+}M(U)$ of
$J$-self-adjoint extensions $A_{M(U)}$ of $A_{\mathrm{sym}}$ and
unitary matrices $U$.
\begin{lelele}\label{k12}
A $J$-self-adjoint extension $A_{M(U)}$ defined by (\ref{e55}) and
(\ref{e6}) is self-adjoint if and only if $q=0$.
\end{lelele}

\emph{Proof.} According to Proposition \ref{t2}, a $J$-self-adjoint
operator $A_{M(U)}$ is self-adjoint if and only if $M(U)$ is also a
hypermaximal neutral subspace in the Krein space $(\mathfrak{M},
[\cdot, \cdot]_{Z})$.

By (\ref{ss1}) the fundamental decomposition of $(\mathfrak{M},
[\cdot, \cdot]_{Z})$ has the form
\begin{equation}\label{as16}\fl
\mathfrak{M}={\mathfrak{N}}_{-i}\oplus{\mathfrak{N}}_{i}, \qquad
{\mathfrak{N}}_{-i}=<e_{-+}, e_{--}>, \quad
{\mathfrak{N}}_{i}=<e_{++}, e_{+-}>,
\end{equation}
where ${\mathfrak{N}}_{-i}$  and ${\mathfrak{N}}_{i}$ are,
respectively, negative and positive subspaces. Taking (\ref{as16})
into account, we derive from (\ref{e6}) that $M(U)$ is a
hypermaximal neutral subspace of $(\mathfrak{M}, [\cdot,
\cdot]_{Z})$ if and only if $q=0$. \rule{2mm}{2mm}

\begin{lelele}\label{k15}
A $J$-self-adjoint extension $A_{M(U)}$ does not have the property
of ${\mathcal{C}}$-symmetry if $r=0$.
\end{lelele}
{\textit Proof.} If $r=0$, then
$d_1=e_{++}+e^{i(\phi+\gamma)}e_{+-}\in{M(U)}\cap\mathfrak{N}_{i}$
(on the basis of (\ref{as16})). In that case $A_{M(U)}d_1=id_1$ by
(\ref{e55}). Therefore $A_{M(U)}$ has a non-real spectrum and there
are no ${\mathcal{C}}$-symmetries for $A_{M(U)}$ (see Corollary
\ref{c23}). \rule{2mm}{2mm}

\noindent {\bf Remark 3.2} Lemmas \ref{k12},  \ref{k15} and the
constraint $q^2+r^2=1$ in \rf{eee6} show that there should exist a
critical angle $\sigma_c\in (0,2\pi)$ in $q=\sin(\sg)$, \
$r=\cos(\sg)$ where the $\cC-$symmetry relation $A_{M(U)}\cC=\cC
A_{M(U)}$ breaks down\footnote{These critical configurations will be
analyzed in a separate paper.}.

\subsection{Family of $C_{\theta,\omega}$-symmetries.}
Let $R$ be a fundamental symmetry in $\mathfrak{H}$ (i.e., $R^2=I$
and $R=R^*$) such that
\begin{equation}\label{as11}
A_{\mathrm{sym}}R=RA_{\mathrm{sym}}, \quad \mbox{and} \quad JR=-RJ.
\end{equation}

The first identity in (\ref{as11}) means that the subspaces
$\mathfrak{N}_{\pm{i}}$  reduce $R$ and the restriction
$R\upharpoonright\mathfrak{M}$ is a fundamental symmetry in the
Hilbert space $\mathfrak{M}$. The second identity and the definition
of the elements $\{e_{\pm\pm}\}$ imply
\begin{equation}\label{as12}\fl
Re_{++}=e_{+-}, \quad Re_{+-}=e_{++}, \quad  Re_{--}=e_{-+}, \quad
Re_{-+}=e_{--}.
\end{equation}

Furthermore, the relation $JR=-RJ$ enables one to state that the
operator
\begin{equation}\label{e34}
R_\omega=Re^{i{\omega}J}=e^{-i{\omega}J/2}Re^{i{\omega}J/2}, \qquad
\omega\in[0, 2\pi).
\end{equation}
is an involution ($R^2_\omega=I$,\ $R_\omega=R_\omega^*$) in
$\mathfrak{H}$ and $JR_\omega=-R_\omega{J}$. It follows from
(\ref{ss1}), (\ref{as12}), and (\ref{e34}) that
\begin{equation}\label{e4}\fl
R_{\omega}e_{++}=e^{i\omega}e_{+-}, \
R_{\omega}e_{+-}=e^{-i\omega}e_{++}, \
R_{\omega}e_{--}=e^{-i\omega}e_{-+}, \
R_{\omega}e_{-+}=e^{i\omega}e_{--}.
\end{equation}

Let us consider the collection of operators
$$
T_{\theta,\omega}=\frac{1-\theta}{1+\theta}R_\omega, \qquad
\theta>0, \quad \omega\in[0, 2\pi).
$$

Obviously, $T_{\theta, \omega}$ is self-adjoint in $\mathfrak{H}$,
$JT_{\theta, \omega}=-T_{\theta, \omega}J$, and $\|T_{\theta,
\omega}\|<1$. By (\ref{d4}) and (\ref{e3}), $T_{\theta, \omega}$ is
the operator of transition from (\ref{d1}) to the decomposition
\begin{equation}\label{dd4}\fl
{\mathfrak{H}}={\mathfrak L}^{\theta,\omega}_+[\oplus]_J{\mathfrak
L}^{\theta, \omega}_-, \qquad {\mathfrak L}^{\theta,
\omega}_+=(I+T_{\theta, \omega})\mathfrak{H}_+, \qquad {\mathfrak
L}^{\theta, \omega}_-=(I+T_{\theta, \omega})\mathfrak{H}_-.
\end{equation}

Let us introduce the notation
$$
\alpha_\theta=\frac{1}{2}(\theta+{\theta}^{-1})=\cosh(\chi) \quad
\mbox{and} \quad
\beta_\theta=\frac{1}{2}(\theta-{\theta}^{-1})=\sinh(\chi), \quad
\theta=e^\chi
$$
so that $\alpha_\theta^2-\beta_\theta^2=1$. Due to (\ref{e2}) the
operator ${\mathcal C}_{\theta, \omega}$ associated with (\ref{dd4})
has the form
\begin{equation}\label{e13}
{\mathcal C}_{\theta,\omega}=(I+T_{\theta, \omega})(I-T_{\theta,
\omega})^{-1}J=[\alpha_\theta{I}-\beta_\theta{R_\omega}]J=e^{-\chi
R_\omega}J.
\end{equation}
In particular $ {\mathcal C}_{1,\omega}=J,\
\forall\omega\in[0,2\pi).$ Moreover, due to \rf{aaa3} one has
$Q=-\chi R_\omega$.

By Theorem \ref{t1} and (\ref{e13}) the decomposition (\ref{dd4})
can be rewritten as
\begin{equation}\label{dede4}\fl
{\mathfrak{H}}={\mathfrak L}_+^{\theta,\omega}\oplus_{\mathcal
C}{\mathfrak L}_-^{\theta, \omega}, \quad {\mathfrak L}_+^{\theta,
\omega}=\frac{1}{2}(I+{\mathcal{C}}_{\theta,\omega})\mathfrak{H},
\quad {\mathfrak L}_-^{\theta,
\omega}=\frac{1}{2}(I-{\mathcal{C}}_{\theta,\omega})\mathfrak{H}.
\end{equation}
(The formulas (\ref{dd4}) and (\ref{dede4}) determine the same
decomposition of $\mathfrak{H}$; the first formula emphasizes the
$J$-orthogonality of ${\mathfrak L}_{\pm}^{\theta,\omega}$, the
second one illustrates the orthogonality of ${\mathfrak
L}_{\pm}^{\theta,\omega}$ with respect to the inner product $(\cdot,
\cdot)_{\mathcal{C}}$.)

\begin{lelele}\label{l34}
The following relations hold:
\begin{equation}\label{new14}
{\mathcal C}_{\theta, \omega}^2=I, \qquad {\mathcal C}_{\theta,
\omega}^*={\mathcal C}_{1/\theta, \omega} \qquad J{\mathcal
C}_{\theta_1, \omega}{\mathcal C}_{\theta_2, \omega}={\mathcal
C}_{{\theta_2}/{\theta_1}, \omega}.
\end{equation}
Furthermore, $\|{\mathcal C}_{\theta, \omega}\|=\theta$ if
$\theta\geq1$ and $\|{\mathcal C}_{\theta, \omega}\|=1/\theta$ if
$\theta<1$.
\end{lelele}

{\it Proof.} The relations (\ref{new14}) immediately follow from
(\ref{e34}) and (\ref{e13}). By virtue of (\ref{aaa3}), ${\mathcal
C}_{\theta, \omega}J=e^{-\chi R_{\omega}}$ with $R_\omega$ a bounded
self-adjoint operator.  According to (\ref{e13}),
\begin{equation}\label{ad1}\fl
(e^{-\chi
R_{\omega}}x,x)=\alpha_\theta\|x\|^2-\beta_\theta(R_\omega{x},x)\leq(\alpha_\theta+|\beta_\theta|)\|x\|^2,
\quad \forall{x}\in\mathfrak{H}.
\end{equation}
Obviously, (\ref{ad1}) turns out to be identity for any
$x\in\ker(R_\omega+\sign(\beta_\theta){I})$. Therefore, $\|{\mathcal
C}_{\theta, \omega}\|=\|e^{-\chi
R_{\omega}}\|=\alpha_\theta+|\beta_\theta|$ since $e^{-\chi
R_{\omega}}$ is a positive self-adjoint operator. Recalling the
definition of $\alpha_\theta$ and $\beta_\theta$ we complete the
proof of the Lemma. \rule{2mm}{2mm}

\subsection{The description of $J$-self-adjoint extensions with
${\mathcal{C}}_{\theta,\omega}$-symmetries.} Let $A_{M(U)}$ be a
$J$-self-adjoint extension of $A_{\mathrm{sym}}$ defined by
(\ref{e55}) and (\ref{e6}).

\begin{lelele}\label{k14}
A $J$-self-adjoint extension $A_{M(U)}$ has
${\mathcal{C}}_{1,\omega}$-symmetry if and only if $q=0$ (or,
equivalently, $A_{M(U)}$ is self-adjoint).
\end{lelele}
\emph{Proof.} A $J$-self-adjoint extension $A_{M(U)}$ has
${\mathcal{C}}_{1,\omega}$-symmetry $\iff{A_{M(U)}}J=J{A_{M(U)}}$.
Comparing this with the relation $A_{M(U)}^*J=JA_{M(U)}$ (since
$A_{M(U)}$ is $J$-self-adjoint) one derives that
$A_{M(U)}^*=A_{M(U)}$. Applying now Lemma \ref{k12} we complete the
proof. \rule{2mm}{2mm}

\begin{ddd}\label{ddd-y} Let
$\Upsilon$ denote the collection of all $J$-self-adjoint extensions
$A_{M(U)}$ having ${\mathcal{C}}_{\theta,\omega}$-symmetry \emph{for
any choice} of $\theta$ and $\omega$: $$\Upsilon=\{A_{M(U)}:\
A_{M(U)}\cC=\cC A_{M(U)},\ \forall \theta \in (0,\infty) \cup
\forall \omega\in [0,2\pi)\}.$$
\end{ddd}
\noindent In analogy to Lie algebra theory \cite{Kn} it appears
natural to call $\Upsilon$ the \emph{extension center}.

Obviously, an operator $A_{M(U)}\in\Upsilon$ is self-adjoint (since
$A_{M(U)}$ has ${\mathcal{C}}_{1,\omega}$-symmetry) and it has a
special structure closely related to the properties of
$A_{\mathrm{sym}}$. One of the possible ways to describe this
structure deals with the concept of supersymmetry (SUSY).

Let $H$ and $\cQ$ be self-adjoint operators in $\mathfrak{H}$.
Following \cite{CFKS} we will say that the system $(H,J,\cQ)$
possesses \emph{supersymmetry} if $H=\cQ^2\geq{0}$ and $J\cQ=-\cQ
J$.
\begin{ppp}\label{new2}
Let $A_{M(U)}$ be a $J$-self-adjoint extension of
$A_{\mathrm{sym}}$. The following statements are equivalent:

$(i)$ \ \ $A_{M(U)}$ belongs to $\Upsilon$;

$(ii)$ \ $A_{M(U)}J=JA_{M(U)}$ \ and  \ $A_{M(U)}R=RA_{M(U)}$;

$(iii)$ the system $(A_{M(U)}^2,J,RA_{M(U)})$ has supersymmetry.
\end{ppp}

\emph{Proof.} It follows from (\ref{e34}) and (\ref{e13}) that
$A_{M(U)}\in\Upsilon$ if and only if $JA_{M(U)}=A_{M(U)}J$ and
$RA_{M(U)}=A_{M(U)}R$. So, $(i)\iff(ii)$. The latter relation and
$JR=-RJ$ mean that $\cQ=RA_{M(U)}$ is self-adjoint and $J\cQ=-\cQ
J$. Since $H=(RA_{M(U)})^2=A_{M(U)}^2\geq{0}$ the system
$(A_{M(U)}^2,J,RA_{M(U)})$ has supersymmetry.

Conversely, if $(A_{M(U)}^2,J,RA_{M(U)})$ has supersymmetry, then
$JRA_{M(U)}=-RA_{M(U)}J$ or $JA_{M(U)}=A_{M(U)}J$. Therefore, the
$J$-self-adjoint operator $A_{M(U)}$ is also self-adjoint. In that
case the self-adjointness of $RA_{M(U)}$ gives:
$RA_{M(U)}=(RA_{M(U)})^*=A_{M(U)}R$. So, $A_{M(U)}$ commutes with
$J$ and $R$. Hence, $(ii)\iff(iii)$. \rule{2mm}{2mm}

The next statement gives the description of extension center
elements $A_{M(U)}\in\Upsilon$ in terms of entries of $U$ (see
(\ref{eee6})).

\begin{ppp}\label{k17}
$A_{M(U)}\in\Upsilon \ \iff \ q=0$ and $\phi\in\{\frac{\pi}{2},
\frac{3\pi}{2}\}$.
\end{ppp}
\emph{Proof.} Let $A_{M(U)}\in\Upsilon$.
 Since $A_{\mathrm{sym}}$ commutes with $J$ and $R$, assertion $(ii)$ of
Proposition \ref{new2} can be rewritten as: $J{M(U)}={M(U)}$ and
$R{M(U)}={M(U)}$.

It follows from (\ref{ss1}) and the description (\ref{e6}) of
${M(U)}$ that $J{M(U)}={M(U)}$ if and only if
$$
\begin{array}{c}
Jd_1=e_{++}-qe^{i(\phi+\gamma)}e_{+-}+re^{i(\phi+\xi)}e_{-+}\in{M(U)},
\\
Jd_2=-e_{--}+re^{i(\phi-\xi)}e_{+-}+qe^{i(\phi-\gamma)}e_{-+}\in{M(U)}.
\end{array}
$$
This is possible if and only if $q=0$ (since $\{e_{\pm\pm}\}$ are
orthonormal and $d_i$ have the form (\ref{aaa5})).

A similar reasoning for $R{M(U)}={M(U)}$ with the use of
(\ref{as12}) gives
$$
\begin{array}{c}
Rd_1=R(e_{++}+re^{i(\phi+\xi)}e_{-+})=re^{i(\phi+\xi)}(e_{--}+re^{-i(\phi+\xi)}e_{+-})\in{M(U)}
\\
Rd_2=R(e_{--}-re^{i(\phi-\xi)}e_{+-})=-re^{i(\phi-\xi)}(e_{++}-re^{i(-\phi+\xi)}e_{-+})\in{M(U)},
\end{array}
$$
where $r^2=1$. Obviously the latter relations are satisfied if and
only if $e^{-i\phi}=-e^{i\phi}$. This is possible when
$\phi=\frac{\pi}{2}$ or  $\phi=\frac{3\pi}{2}$ . Proposition
\ref{k17} is proved. \rule{2mm}{2mm}

\begin{ttt}\label{t15}
Let $A_{M(U)}$ be a $J$-self-adjoint extension of $A_{\mathrm{sym}}$
and $A_{M(U)}\not=A_{M(U)}^*$ (i.e. $A_{M(U)}$ is not a self-adjoint
operator). Then $A_{M(U)}$ has
${\mathcal{C}}_{\theta,\omega}$-symmetry if and only if
\begin{equation}\label{es21}
0<|q|<|\cos\phi|.
\end{equation}
In that case $\omega=\gamma$ and $\theta$ is determined by the
relation $q=\frac{\theta^{-1}-\theta}{\theta^{-1}+\theta}\cos\phi.$
\end{ttt}

\emph{Proof.} Since $A_{\mathrm{sym}}$ commutes with $J$ and $R$ it
commutes with $R_\omega$ defined by (\ref{e34}). This gives
$A_{\mathrm{sym}}{\mathcal{C}}_{\theta,\omega}={\mathcal{C}}_{\theta,\omega}A_{\mathrm{sym}}$
(since ${\mathcal{C}}_{\theta,\omega}$ has the form (\ref{e13})).
Employing Theorem \ref{p1} one concludes that the property of
${{\mathcal{C}}}_{{\theta},\omega}$-symmetry for $A_{M(U)}$ is
equivalent to the relation
${{\mathcal{C}}}_{{\theta},\omega}M(U)=M(U)$.
 By (\ref{e6}), ${\mathcal{C}}_{\theta,\omega}M(U)=M(U)\iff$
${\mathcal{C}}_{\theta,\omega}d_1\in{M(U)}$ and
${\mathcal{C}}_{\theta,\omega}d_2\in{M(U)}$, where $d_i$ have the
form (\ref{aaa5}).

It follows from (\ref{ss1}), (\ref{e4}) and (\ref{e13}) that
\begin{eqnarray}\label{e2223}\fl
 {\mathcal{C}}_{\theta,\omega}d_1  = (\alpha_\theta+\beta_\theta{q}e^{i(\gamma+\phi-\omega)})e_{++}
& & \nonumber
\\ \fl
-(\beta_\theta{e^{i\omega}}+\alpha_\theta{q}e^{i(\gamma+\phi-\omega)})e_{+-}+\alpha_{\theta}re^{i(\xi+\phi)}e_{-+}-\beta_{\theta}re^{i(\xi+\phi+\omega)}e_{--}.&
&
\end{eqnarray}

Taking the definition of $d_1$ and the first and the last terms in
(\ref{e2223}) into account one concludes that
${\mathcal{C}}_{\theta,\omega}d_1\in{M(U)}\iff{{\mathcal{C}}_{\theta,\omega}d_1}=k_1d_1+k_2d_2$,
where $k_1=\alpha_\theta+\beta_\theta{q}e^{i(\gamma+\phi-\omega)}$
and $k_2=-\beta_{\theta}re^{i(\xi+\phi+\omega)}$. This is possible
if and only if the following equalities are satisfied:
\begin{eqnarray}\label{e223}
\beta_\theta{qr}e^{i(\gamma+\xi+2\phi-\omega)}=\beta_\theta{qr}e^{i(-\gamma+\xi+2\phi+\omega)}
& & \nonumber \\
\beta_\theta{q}^2e^{i(2\gamma+2\phi-\omega)}+2\alpha_\theta{q}e^{i(\gamma+\phi)}+\beta_\theta{e^{i\omega}}(r^2e^{2i\phi}+1)=0.
\end{eqnarray}

A similar reasoning for
${\mathcal{C}}_{\theta,\omega}d_2=\widetilde{k}_1d_1+\widetilde{k}_2d_2$
with $\widetilde{k}_1=-\beta_{\theta}re^{i(-\xi+\phi-\omega)}$ and
$\widetilde{k}_2=-\alpha_\theta-\beta_{\theta}qe^{i(-\gamma+\phi+\omega)}$
implies
\begin{eqnarray}\label{e224}
-\beta_\theta{qr}e^{i(\gamma-\xi+2\phi-\omega)}=-\beta_\theta{qr}e^{i(-\gamma-\xi+2\phi+\omega)}
& &
\nonumber \\
\beta_\theta{q}^2e^{i(-2\gamma+2\phi+\omega)}+2\alpha_\theta{q}e^{i(-\gamma+\phi)}+\beta_\theta{e^{-i\omega}}(r^2e^{2i\phi}+1)=0.
\end{eqnarray}
Therefore, $A_{M(U)}$ has ${\mathcal{C}}_{\theta,\omega}$-symmetry
if and only if relations (\ref{e223}) and (\ref{e224}) hold.

Let $A_{M(U)}$ have ${\mathcal{C}}_{\theta,\omega}$-symmetry and
$A_{M(U)}\not=A_{M(U)}^*$. Then  $\theta\not=1$ (otherwise,
$A_{M(U)}$ turns out to be self-adjoint). Further, $q\not=0$ (by
Lemma \ref{k12}), $r\not=0$ (by Lemma \ref{k15}), and
$\beta_\theta\not=0$ (since $\theta\not=1$). Taking these facts into
account we derive from (\ref{e223}) and (\ref{e224}) that
${\mathcal{C}}_{\theta\not=1,\omega}M(U)=M(U)$ if and only if
\begin{equation}\label{as2}\fl
\omega=\gamma \quad \mbox{and} \quad
\beta_\theta{q}^2e^{i(2\phi+\omega)}+2\alpha_\theta{q}e^{i(\omega+\phi)}+\beta_\theta{e^{i\omega}}(r^2e^{2i\phi}+1)=0.
\end{equation}
Since $q^2+r^2=1$ (by (\ref{eee6})) the second relation in
(\ref{as2}) can be rewritten as
\begin{equation}\label{as3}
q=-\frac{\beta_\theta}{\alpha_\theta}\left[\frac{e^{i\phi}+e^{-i\phi}}{2}\right]=
\frac{\theta^{-1}-\theta}{\theta^{-1}+\theta}\cos\phi.
\end{equation}
Since $\theta\not=1$, the relation (\ref{as3}) implies inequality
(\ref{es21}).

 Conversely, let the parameters $\phi$ and $q$ of the unitary matrix $U$
 (see (\ref{eee6})) satisfy (\ref{es21}). Then the corresponding $J$-self-adjoint extension
$A_{M(U)}$ does not have ${\mathcal{C}}_{1,\omega}$-symmetry and
hence $A_{M(U)}$ is not a self-adjoint operator.

The condition (\ref{es21}) allows one to choose a parameter $\theta$
$(\theta\not=1)$ in such a way that (\ref{as3}) holds. Finally
setting $\omega=\gamma$, we satisfy the relations (\ref{as2}). This
means that $A_{M(U)}$ has ${\mathcal{C}}_{\theta,\omega}$-symmetry
for such a choice of $\omega$ and $\theta$. Theorem \ref{t15} is
proved.\rule{2mm}{2mm}

\begin{ttt}\label{k18}
A $J$-self-adjoint extension $A_{M(U)}$ of $A_{\mathrm{sym}}$ has
${\mathcal{C}}_{\theta,\omega}$-symmetry if and only if the matrix
$U$ takes the form
\begin{equation}\label{as34}\fl
U=U(\theta,\omega,\psi,\xi)=\frac{e^{i\phi}}{\alpha_\theta}\left(\begin{array}{cc}
-\beta_\theta\cos\phi{e}^{i\omega} & \sqrt{1+\beta^2_\theta\sin^2\phi}{e}^{i\xi} \vspace{2mm} \\
-\sqrt{1+\beta^2_\theta\sin^2\phi}e^{-i\xi} &
-\beta_\theta\cos\phi{e}^{-i\omega}\end{array}\right),
\end{equation}
where $\phi, \xi\in[0,2\pi)$.
\end{ttt}
\emph{Proof.} Let us consider the case $\theta\not=1$ and
$\phi\not\in\{\frac{\pi}{2}, \frac{3\pi}{2}\}$. Then (\ref{as34}) is
a particular case of the general representation of unitary matrices
(\ref{eee6}) with $q=-\frac{\beta_\theta}{\alpha_\theta}\cos\phi$
that satisfies (\ref{es21}). This means that the $J$-self-adjoint
operator $A_{M(U)}$ has ${\mathcal{C}}_{\theta,\omega}$-symmetry (by
Theorem \ref{t15}).

Conversely, let $U=\|u_{ij}\|$ be determined by (\ref{eee6}) with
$\phi\not\in\{\frac{\pi}{2}, \frac{3\pi}{2}\}$ and let the
corresponding $J$-self-adjoint extension $A_{M(U)}$ have
${\mathcal{C}}_{\theta\not=1,\omega}$-symmetry. Due to (\ref{as2})
and (\ref{as3}),
$u_{11}=qe^{i(\phi+\gamma)}=-\frac{\beta_\theta}{\alpha_\theta}\cos\phi{e}^{i(\phi+\omega)}$.
But then
$u_{22}=-\frac{\beta_\theta}{\alpha_\theta}\cos\phi{e}^{i(\phi-\omega)}$
by (\ref{eee6}). Similarly,
\begin{eqnarray*}
u_{12}&=&re^{i(\phi+\xi)}=\sqrt{1-q^2}e^{i(\phi+\xi)}=\frac{1}{\alpha_\theta}\sqrt{\alpha^2_\theta-\beta^2_\theta\cos^2\phi}{e}^{i(\phi+\xi)}\nn\\
&=&\frac{1}{\alpha_\theta}\sqrt{1+\beta^2_\theta\sin^2\phi}{e}^{i(\phi+\xi)}.
\end{eqnarray*}
and
$u_{22}=-re^{i(\phi-\xi)}=-{\frac{1}{\alpha_\theta}\sqrt{1+\beta^2_\theta\sin^2\phi}}e^{i(\phi-\xi)}.
$ Hence, the matrix $U$ is determined by (\ref{as34}).

Let $\theta=1$ and let $\phi$ be arbitrary. By Lemma \ref{k14}
$J$-self-adjoint extension $A_{M(U)}$ with
${\mathcal{C}}_{1,\omega}$-symmetry is self-adjoint and $q=0$. In
that case the representation (\ref{eee6}) of $U$ coincides with
(\ref{as34}).

Let $\theta\not=1$  and $\phi\in\{\frac{\pi}{2}, \frac{3\pi}{2}\}$.
It follows from Theorem \ref{t15} that $A_{M(U)}$ has to be
self-adjoint (otherwise, the inequality (\ref{es21}) must be
satisfied, what is impossible since $\phi\in\{\frac{\pi}{2},
\frac{3\pi}{2}\}$). Hence, $q=0$ (by Lemma \ref{k12}) and the
representation (\ref{eee6}) of $U$ coincides with (\ref{as34}).
Theorem \ref{k18} is proved. \rule{2mm}{2mm}

\subsection{Completeness of the ${\mathcal{C}}_{\theta,\omega}$-symmetry family}
As was mentioned above (see the proof of Theorem \ref{t15}), an
arbitrary operator ${\mathcal{C}}_{\theta,\omega}$ from the
2-parameter set $\{{\mathcal{C}}_{\theta,\omega}\}$ commutes with
$A_{\mathrm{sym}}$. We are going to show that, in a certain sense,
this family is complete in the set of ${\mathcal{C}}$-symmetries
commuting with $A_{\mathrm{sym}}$. Precisely, we show that an
arbitrary $J$-self-adjoint extension
$A_{M(U)}{\supset}A_{\mathrm{sym}}$ having the property of
${\mathcal{C}}$-symmetry, where ${\mathcal{C}}$ commutes with
$A_{\mathrm{sym}}$, possesses a
${\mathcal{C}}_{\theta,\omega}$-symmetry for some choice of $\theta$
and $\omega$. From this point of view, the family
${\mathcal{C}}_{\theta,\omega}$ allows for an adequate description
of the set of ${\mathcal{C}}$-symmetries commuting with
$A_{\mathrm{sym}}$.

Our proof below requires the existence of at least one real point
$\lambda$ of regular type for the initial symmetric operator
$A_{\mathrm{sym}}$, which is defined in the standard manner as:
$\lambda\in\mathbb{R}$ is a point of regular type of
$A_{\mathrm{sym}}$ if there exists a number $k=k(\lambda)>0$ such
that $\|(A_{\mathrm{sym}}-\lambda{I})u\|\geq{k}\|u\|$, \
$\forall{u}\in\mathcal{D}(A_{\mathrm{sym}})$. This condition is not
restrictive because it is satisfied for any symmetric operator
$A_{\mathrm{sym}}$ having at least one self-adjoint extension $A$
with spectrum $\sigma(A)$ which is not covering the whole real line
$\mathbb{R}$ (i.e., $\sigma(A)\not=\mathbb{R}$).
\begin{ttt}\label{p14}
Let a symmetric operator $A_{\mathrm{sym}}$ with deficiency indices
$<2,2>$ have at least one real point $\lambda$ of regular type and
let a $J$-self-adjoint extension $A_{M(U)}\supset{A}_{\mathrm{sym}}$
have the property of ${\mathcal{C}}$-symmetry, where ${\mathcal{C}}$
commutes with $A_{\mathrm{sym}}$. Then $A_{M(U)}$ also has the
property of ${\mathcal{C}}_{\theta,\omega}$-symmetry for a certain
choice of $\theta$ and $\omega$.
\end{ttt}

The proof of Theorem \ref{p14} is based on the following auxiliary
result.
\begin{lelele}\label{p11}
Let  $A_{\mathrm{sym}}$ satisfy the conditions of Theorem \ref{p14}
and let $A_{\mathrm{sym}}\mathcal{C}=\mathcal{C}A_{\mathrm{sym}}$,
where $\mathcal{C}$ is a bounded linear operator in $\mathfrak{H}$
with the properties: ${\mathcal{C}}^2=I$ and $J\mathcal{C}>0$. Then
the restrictions of ${\mathcal{C}}$ onto
$\mathfrak{M}=\mathfrak{N}_{i}\dot{+}\mathfrak{N}_{-i}$ coincide
with the restriction of $\mathcal{C}_{\theta,\omega}$ for a certain
choice of $\theta$ and $\omega$, i.e.,
$\mathcal{C}\upharpoonright\mathfrak{M}=\mathcal{C}_{\theta,
\omega}\upharpoonright\mathfrak{M}$.
\end{lelele}

\emph{Proof of Theorem \ref{p14}.} Let $J$-self-adjoint extension
$A_{M(U)}\supset{A}_{\mathrm{sym}}$ have the property of
${\mathcal{C}}$-symmetry, where ${\mathcal{C}}$ commutes with
$A_{\mathrm{sym}}$. Then ${\mathcal{C}}M(U)=M(U)$ by Theorem
\ref{p1}. Since $M(U)\subset{\mathfrak{M}}$, the last equality is
equivalent to ${\mathcal{C}}_{\theta,\omega}M(U)=M(U)$ for a certain
choice of $\theta$ and $\omega$ by Lemma \ref{p11}. Using Theorem
\ref{p1} again one derives the property of
${\mathcal{C}}_{\theta,\omega}$-symmetry for $A_{M(U)}$.
\rule{2mm}{2mm}

\emph{Proof of Lemma \ref{p11}.}
It follows from the proof of
Theorem \ref{p1} that
$\mathcal{C}\mathfrak{N}_{\pm{i}}=\mathfrak{N}_{\pm{i}}$. Therefore,
$\mathcal{C}$ has the block structure $
\cC=\left(%
\begin{array}{cc}
  \cC_+ & 0 \\
  0 & \cC_- \\
\end{array}%
\right) \
(\cC_{\pm}:=\mathcal{C}\upharpoonright\mathfrak{N}_{\pm{i}})$ with
respect to the decomposition
$\mathfrak{M}=\mathfrak{N}_{i}\dot{+}\mathfrak{N}_{-i}$.

Let us fix $\mathfrak{N}_{i}$ and consider the Pauli matrices
\begin{equation}\label{pauli}
\sigma_1=\left(\begin{array}{cc} 0  & 1 \\
1 & 0
\end{array}\right), \quad \sigma_2=\left(\begin{array}{cc} 0  & -i \\
i & 0
\end{array}\right), \quad \sigma_3=\left(\begin{array}{cc} 1  & 0 \\
0 & -1
\end{array}\right).
\end{equation}

Since $\mathfrak{N}_{i}=<e_{++},e_{+-}>$, formulas (\ref{ss1}) and
(\ref{as12}) imply that $J=\sg_3$ and $R=\sg_1$ with respect to the
basis $\{e_{++},e_{+-}\}$.

The conditions ${\mathcal{C}}^2=I$ and $J\mathcal{C}>0$ imposed on
${\mathcal{C}}$ in Lemma \ref{p11} together with \rf{e13} enable one
to represent ${\mathcal{C}}$ as follows:  $\cC =e^{-\chi
R_\omega}J$, where due to \rf{e34}  $R_\omega J=-JR_\omega$,
$R_\omega=R_\omega^*$ and $R_\omega^2=I$. Obviously, the same
relation must hold for the $2\times 2$ matrix $\cC_+$, i.e. $\cC_+=
e^{-\chi_+ R_{\omega_1}}\sg_3$ with
\begin{equation}\label{s1}
R_{\omega_1} =Re^{i\omega
J}=\cos(\omega_1)\sg_1+\sin(\omega_1)\sg_2.
\end{equation}
{}From the relation $R_{\omega_1}^2=I_2$ it follows
\ba{8} e^{-\chi_+R_{\omega_1}}&=&\cosh (\chi_+)I_2- \sinh(\chi_+)R_{\omega_1}.
\ea
Identifying $\a_{\theta_1}=\cosh (\chi_+)$, $\b_{\theta_1}=\sinh
(\chi_+)$ and using \rf{s1} we get for $\cC_+=
e^{-\chi_+R_{\omega_1}}\sg_3$ the explicit representation
\ba{10}
\cC_+&=&\left(%
\begin{array}{cc}
  \a_{\theta_1} & \b_{\theta_1}  e^{-i\om_1} \\
  -\b_{\theta_1}  e^{i\om_1} & -\a_{\theta_1} \\
\end{array}%
\right) \ea with respect to the basis $\{e_{++},e_{+-}\}$.

On the other hand, relations (\ref{ss1}), (\ref{e4}), and
(\ref{e13}) mean that the operator ${\mathcal{C}}_{\theta_1,
\omega_1}\upharpoonright{\mathfrak{N}_{i}}$ has the same matrix
representation (\ref{10}) with respect to $\{e_{++},e_{+-}\}$.
Therefore,
$\cC_+=\mathcal{C}\upharpoonright\mathfrak{N}_{i}=\mathcal{C}_{\theta_1,
\omega_1}\upharpoonright\mathfrak{N}_{i}$.

It should be noted that parameters $\theta_1$, $\omega_1$ in
(\ref{10}) are not determined uniquely and that the pairs $\theta_1,
\omega_1$ and $1/\theta_1, \omega_1-\pi$ define the same matrix
$\cC_+$. In what follows, without loss of generality we will suppose
$\theta_1\geq1$.

Arguing similarly one derives \ba{14}
\cC_-&=&\left(%
\begin{array}{cc}
  \a_{\theta_2} & \b_{\theta_2}  e^{-i\om_2} \\
  -\b_{\theta_2}  e^{i\om_2} & -\a_{\theta_2} \\
\end{array}%
\right), \qquad \theta_2\geq1 \ea with respect to the basis
$\{e_{-+},e_{--}\}$ of $\mathfrak{N}_{-i}$ and
$\cC_-=\mathcal{C}_{\theta_2,
\omega_2}\upharpoonright\mathfrak{N}_{-i}$.

Let us show that $\theta_1=\theta_2$ and $\omega_1=\omega_2$. To
prove this we fix a real point $\lambda$ of regular type of
$A_{\mathrm{sym}}$ and consider an operator
$$
A(u+x_\lambda)=A_{\mathrm{sym}}u+\lambda{x_\lambda}, \quad
{\mathcal{D}}(A)={\mathcal{D}}(A_{\mathrm{sym}})\dot{+}{\mathfrak{N}}_{-\lambda}
\quad
({\mathfrak{N}}_{-\lambda}=\mathfrak{H}\ominus\mathcal{R}(A_{\mathrm{sym}}-{\lambda}I)).
$$
Since the real point $\lambda$ is of regular type, the operator $A$
is a self-adjoint extension of $A_{\mathrm{sym}}$. Furthermore, the
commutativity of $A_{\mathrm{sym}}$ with the family
$\{{\mathcal{C}}_{\theta,\omega}\}$ gives
${\mathcal{C}}_{\theta,\omega}{\mathfrak{N}}_{-\lambda}={\mathfrak{N}}_{-\lambda}$.
Therefore,
$A{\mathcal{C}}_{\theta,\omega}={\mathcal{C}}_{\theta,\omega}A$ for
any choice of $\omega$ and $\theta$. Thus $A=A_{M(U)}\in\Upsilon$.
In that case Proposition \ref{k17} allows one to simplify the
general description ${M(U)}$ given by (\ref{e6}) and (\ref{aaa5}) as
follows:
\begin{equation}\label{aaa10}\fl
M(U)=<d_1,d_2>, \quad d_1=e_{++}+ie^{i\xi}e_{-+}, \quad
d_2=e_{--}-ie^{-i\xi}e_{+-}.
\end{equation}

Turning to the original operator $\mathcal{C}$ we deduce from the
proof of Theorem \ref{p1} that
$\mathcal{C}^*A_{\mathrm{sym}}=A_{\mathrm{sym}}\mathcal{C}^*$. This
gives
$\mathcal{C}{\mathfrak{N}}_{-\lambda}={\mathfrak{N}}_{-\lambda}$ and
hence, the operator $A=A_{M(U)}$ commutes with $\mathcal{C}$.
Employing Theorem \ref{p1} one derives $\mathcal{C}M(U)=M(U)$, where
$M(U)$ is defined by (\ref{aaa10}). Taking the relations (\ref{10})
and (\ref{14}) into account and arguing as in the proof of Theorem
\ref{t15} we conclude that the equality ${{\mathcal{C}}}M(U)=M(U)$
is equivalent to the relations
\begin{equation}\label{es23}
\alpha_{\theta_1}=\alpha_{\theta_2}, \qquad
\beta_{\theta_1}e^{i\omega_1}=\beta_{\theta_2}e^{i\omega_2}.
\end{equation}

The first relation in (\ref{es23}) gives
$\theta:=\theta_1=\theta_2$. If $\theta=1$, then the second relation
in (\ref{es23}) vanishes. In that case $\mathcal{C}_{1,
\omega_1}=\mathcal{C}_{1, \omega_2}=J$ and the restriction
$\mathcal{C}\upharpoonright\mathfrak{M}$ coincides with $J$. If
$\theta>1$ then $\beta_{\theta}\not=0$ and the second relation in
(\ref{es23}) gives $\omega:=\omega_1=\omega_2$. Hence,
$\mathcal{C}\upharpoonright\mathfrak{M}=\mathcal{C}_{\theta,
\omega}$. Lemma \ref{p11} is proved. \rule{2mm}{2mm}

\noindent {\bf Remark 3.3} Physically, $\cC_{\pm}=\exp[-\chi_\pm
R_{\omega_{1,2}}/2](J\upharpoonright\mathfrak{N}_{\pm{i}})\exp[\chi_\pm
R_{\omega_{1,2}}/2]$ in \rf{10} and \rf{14} are just the
hyperbolically rotated (boosted) versions of the involution
$J\upharpoonright\mathfrak{N}_{\pm{i}}$. The transformation matrices
$\exp[\chi_\pm R_{\omega_{1,2}}/2]$ are elements of the
pseudounitary group $SU(1,1)\cong SO(1,2)\cong SL(2,\RR)$ \cite{Vi}
with $R_\omega=e^{-i\omega J/2} Re^{i\omega J/2}$ in \rf{s1} as Lie
algebra elements conjugate to $R$ under the transformations of the
compact subgroup $U(1)\cong SO(2)\ni e^{i\omega J/2}$.

\subsection{The resolvent formula.}

As was stated above, the operator $A_{\mathrm{sym}}$ commutes with
the family $\{{\mathcal{C}}_{\theta,\omega}\}$. Therefore, with
respect to the decomposition (\ref{dede4}),  $A_{\mathrm{sym}}$ can
be presented as the direct sum:
$A_{\mathrm{sym}}=A_{\mathrm{sym}}^+\dot{+}A_{\mathrm{sym}}^-$ of
the symmetric operators
$A_{\mathrm{sym}}^{\pm}=A_{\mathrm{sym}}\upharpoonright{\mathfrak
L}_{\pm}^{\theta,\omega}$ acting in the subspaces ${\mathfrak
L}_{\pm}^{\theta,\omega}$ of $\mathfrak{H}$.

Obviously, the defect subspaces
$\mathfrak{N}_{{\pm}i}(A_{\mathrm{sym}}^{+})={\mathfrak
L}_{+}^{\theta,\omega}\ominus\mathcal{R}(A_{\mathrm{sym}}^{+}{\pm}iI)$
of $A_{\mathrm{sym}}^{+}$ coincide with
$\mathfrak{N}_{{\pm}i}\cap{\mathfrak L}_{+}^{\theta,\omega}$, where
$\mathfrak{N}_{{\pm}i}$ are the defect subspaces of
$A_{\mathrm{sym}}$ in $\mathfrak{H}$. Taking this fact and formulas
(\ref{e4}) into account it is easy to verify that
$\mathfrak{N}_{i}(A_{\mathrm{sym}}^{+})=<g_i^{+}(\theta)>$ and
$\mathfrak{N}_{-i}(A_{\mathrm{sym}}^{+})=<g_{-i}^{+}(\theta)>$,
where
\begin{equation}\label{dede1}
g_i^{+}(\theta)=(I+\frac{1-\alpha_\theta}{\beta_\theta}R_\omega)e_{++},
\quad
g_{-i}^{+}(\theta)=(I+\frac{1-\alpha_\theta}{\beta_\theta}R_\omega)e_{-+}
\end{equation}

Arguing similarly for $A_{\mathrm{sym}}^{-}$ one derives
$\mathfrak{N}_{i}(A_{\mathrm{sym}}^{-})=<g_i^{-}(\theta)>$ and
\linebreak[4]
$\mathfrak{N}_{-i}(A_{\mathrm{sym}}^{-})=<g_{-i}^{-}(\theta)>$,
where the defect elements
\begin{equation}\label{dede2}
g_i^{-}(\theta)=(I+\frac{1-\alpha_\theta}{\beta_\theta}R_\omega)e_{+-},
\quad
g_{-i}^{-}(\theta)=(I+\frac{1-\alpha_\theta}{\beta_\theta}R_\omega)e_{--}
\end{equation}
belong to ${\mathfrak L}_{-}^{\theta,\omega}$.

The formulas (\ref{dede1}), (\ref{dede2}) were obtained for
$\theta\not=1$. If $\theta=1$, then:  $g_i^{+}(1)=e_{++}, \
g_{-i}^{+}(1)=e_{-+}, \ g_{i}^{-}(1)=e_{+-}, \
g_{-i}^{-}(1)=e_{--}$.

Note that the norms of $g_{\pm{i}}^{\pm}(\theta)$ are equal to
$\sqrt{\alpha_\theta/(\alpha_\theta+1)}$. Indeed, the orthonormality
of $\{e_{\pm\pm}\}$ in $\mathfrak{M}$ and relations (\ref{aaa2}),
(\ref{as16}) imply $\|e_{\pm\pm}\|^2=1/2$. Taking (\ref{e4}) into
account we deduce from (\ref{dede1})
$$
\|g_i^{+}(\theta)\|^2=\|e_{++}\|^2+\left(\frac{1-\alpha_\theta}{\beta_\theta}\right)^2\|e_{+-}\|^2=\frac{\alpha_\theta}{\alpha_\theta+1}.
$$
The other elements $g_{\pm{i}}^{\pm}(\theta)$ are considered by
analogy.

Let us fix an arbitrary extension center element
$A=A_{M(U)}\in\Upsilon$. According to the definition of $\Upsilon$
(subsection 3.4), $A$ is a self-adjoint extension of
$A_{\mathrm{sym}}$ and $A$ is reduced by the decomposition
(\ref{dede4}) for \emph{an arbitrary choice of $\theta$ and
$\omega$}. The collection of unitary matrices $U$ corresponding to
the operators $A_{M(U)}\in\Upsilon$ is described by (\ref{as34})
with $\phi\in\{\frac{\pi}{2}, \frac{3\pi}{2}\}$. This means that,
without loss of generality (multiplying $e_{+-}$ and $e_{-+}$ by
suitable unimodular constants if it is necessarily), one can assume
that the operator $A=A_{M(U)}$ is defined by the matrix
$U=\left(\begin{array}{cc}
 0 & -1 \\
 -1 & 0
 \end{array}\right)$.

Obviously, $A$ is decomposed as $A=A^+\dot{+}A^-$ with respect to
(\ref{dede4}), where $A^{\pm}$ are self-adjoint extensions of the
symmetric operators $A_{\mathrm{sym}}^{\pm}$ acting in the spaces
${\mathfrak L}_{\pm}^{\theta,\omega}$ and having the deficiency
index $<1,1>$ (due to (\ref{dede1}) and (\ref{dede2})). It is easy
to see that for arbitrary $\theta$ and $\omega$
$$
{\mathcal
D}(A^+)={\mathcal{D}}(A_{\mathrm{sym}}^{+})\dot{+}<g_i^+(\theta)-g_{-i}^+(\theta)>,
\ \  {\mathcal
D}(A^-)={\mathcal{D}}(A_{\mathrm{sym}}^{-})\dot{+}<g_i^-(\theta)-g_{-i}^-(\theta)>.
$$

Let  $A_{M(U)}$ be an arbitrary $J$-self-adjoint extension of
$A_{\mathrm{sym}}$ with ${\mathcal{C}}_{\theta,\omega}$-symmetry.
Then the matrix $U$ has the form (\ref{as34}) (by Theorem \ref{k18})
and the operator $A_{M(U)}$ is reduced by the decomposition
(\ref{dede4}) (\emph{for  fixed $\theta$ and $\omega$}). Therefore,
$A_{M(U)}=A_{M(U)}^+\dot{+}A^-_{M(U)}$, where $A_{M(U)}^\pm$ are
intermediate extensions of $A_{\mathrm{sym}}^{\pm}$ in ${\mathfrak
L}_{\pm}^{\theta,\omega}$. A direct calculation shows:
$$
{\mathcal
D}(A_{{M(U)}}^{\pm})={\mathcal{D}}(A_{\mathrm{sym}}^{\pm})\dot{+}<g_i^{\pm}(\theta)+p_{\pm}g_{-i}^{\pm}(\theta)>,
$$
where
\begin{equation}\label{as14}\fl
p_+=e^{i(\xi+\mu)}, \qquad p_-=-e^{i(\xi-\mu)} \qquad
\left(e^{i\mu}:=\frac{\cos\phi+i\alpha_{\theta}\sin\phi}{|\cos\phi+i\alpha_{\theta}\sin\phi|}\right).
\end{equation}

\begin{ttt}\label{te14}
Let $A\in\Upsilon$ and let $A_{M(U)}$ be an arbitrary
$J$-self-adjoint extension of $A_{\mathrm{sym}}$ with
${\mathcal{C}}_{\theta,\omega}$-symmetry (i.e., the matrix $U$ is
determined by (\ref{as34})). Then, for any
$z\in\mathbb{C}\setminus\mathbb{R}$,
\begin{eqnarray*}\fl
\frac{1}{A_{{M(U)}}-z}=\frac{1}{A-z}+\frac{\alpha_\theta(\alpha_\theta+1)}{{\alpha_\theta}\tan\frac{\xi+\mu}{2}-Q(z)}\left(\frac{A+i}{A-z}\cdot,g_i^+(1/\theta)\right)\frac{A-i}{A-z}g_i^+(\theta)
&  & \\ \fl
-\frac{\alpha_\theta(\alpha_\theta+1)}{{\alpha_\theta}\cot\frac{\xi-\mu}{2}+Q(z)}\left(\frac{A+i}{A-z}\cdot,g_i^-(1/\theta)\right)\frac{A-i}{A-z}g_i^-(\theta),
& &
\end{eqnarray*}
where $\mu=\mu(\phi,\theta)$ is determined in (\ref{as14}) and
$Q(z)=2\left(\frac{1+zA}{A-z}e_{++},({\alpha_\theta}I-{\beta_\theta}R_\omega)e_{++}\right).$
\end{ttt}

\emph{Proof.} Let $z\in\mathbb{C}\setminus\mathbb{R}$ be fixed.
Considering $A^+$ and $A_{{M(U)}}^+$ as one-dimensional
perturbations of the symmetric operator $A_{\mathrm{sym}}^{+}$ in
the space ${\mathfrak L}_{+}^{\theta,\omega}$  and repeating the
standard arguments (see, e.g., \cite[pp. 23--28]{AL1}) one derives
the Krein type resolvent formula
\begin{equation}\label{dede5}\fl
\frac{1}{A_{{M(U)}}^+-z}=\frac{1}{A^+-z}+\frac{1}{i\frac{1-p_+}{1+p_+}\frac{\alpha_\theta}{\alpha_\theta+1}-\widetilde{Q}(z)}\left(\frac{A+i}{A-z}\cdot,g_i^+(\theta)\right)\frac{A-i}{A-z}g_i^+(\theta)
\end{equation}
Here the notation $\frac{1}{B-zI}=(B-zI)^{-1}$ is used and $
\widetilde{Q}(z)=\left(\frac{1+zA}{A-z}g_i^+(\theta),g_i^+(\theta)\right)$
is  Krein's $Q$-function \cite{AL1}. Similarly, the formula
\begin{equation}\label{dede6}\fl
\frac{1}{A_{{M(U)}}^--z}=\frac{1}{A^--z}+\frac{1}{i\frac{1-p_-}{1+p_-}\frac{\alpha_\theta}{\alpha_\theta+1}-Q'(z)}\left(\frac{A+i}{A-z}\cdot,g_i^-(\theta)\right)\frac{A-i}{A-z}g_i^-(\theta)
\end{equation}
relates the resolvents of $A^-$ and $A_{{M(U)}}^-$  in ${\mathfrak
L}_{-}^{\theta,\omega}$. Here
$\widetilde{Q}'(z)=\left(\frac{1+zA}{A-z}g_i^-(\theta),g_i^-(\theta)\right)$.

Let us slightly simplify these formulas. First of all,
$$
i\frac{1-p_+}{1+p_+}=\tan\frac{\xi+\mu}{2} \quad  \mbox{and} \quad
i\frac{1-p_-}{1+p_-}=-\cot\frac{\xi-\mu}{2}
$$ due to (\ref{as14}).
Further, it follows from (\ref{dede1}), (\ref{dede2}), and
(\ref{e4}) that
$$
R_\omega{g_i^{-}(\theta)}=(I+\frac{1-\alpha_\theta}{\beta_\theta}R_\omega)R_\omega{e_{+-}}=e^{-i\omega}{g_i^{+}(\theta)}.
$$
Since $A\in\Upsilon$ and therefore, $A$ commutes with $R_{\omega}$
(see the proof of Proposition \ref{new2}) one concludes:
$$
\widetilde{Q}'(z)=\left(R_\omega\frac{1+zA}{A-z}g_i^-(\theta),{R_\omega}g_i^-(\theta)\right)=\left(\frac{1+zA}{A-z}{R_\omega}g_i^-(\theta),{R_\omega}g_i^-(\theta)\right)=\widetilde{Q}(z).
$$
Furthermore, employing (\ref{dede1}), one derives
$$
\widetilde{Q}(z)=\left(\frac{1+zA}{A-z}e_{++},\left(I+\frac{1-\alpha_\theta}{\beta_\theta}R_\omega\right)^2e_{++}\right)=\frac{\alpha_\theta-1}{\beta_\theta^2}Q(z),
$$
where
$Q(z)=2\left(\frac{1+zA}{A-z}e_{++},({\alpha_\theta}I-{\beta_\theta}R_\omega)e_{++}\right).$

Combining (\ref{dede5}), (\ref{dede6}) with the expressions above
and taking into account that the formula
$f=\frac{(I+\mathcal{C}_{\theta,\omega})}{2}f+\frac{(I-\mathcal{C}_{\theta,\omega})}{2}f$
gives the decomposition of an arbitrary element $f\in\mathfrak{H}$
into its ${\mathfrak L}_{\pm}^{\theta,\omega}$-parts, one gets
(after trivial calculations) the following resolvent formula in
$\mathfrak{H}$:
\begin{eqnarray*}\fl
\frac{1}{A_{{M(U)}}-z}&=&\frac{1}{A-z}+
\\ \fl
&&+\frac{\beta_\theta^2}{(\alpha_\theta-1)[{\alpha_\theta}\tan\frac{\xi+\mu}{2}-Q(z)]}\left(\frac{A+i}{A-z}\frac{I+\mathcal{C}_{\theta,\omega}}{2}\cdot,g_i^+(\theta)\right)\frac{A-i}{A-z}g_i^+(\theta)
\\ \fl
&&-\frac{\beta_\theta^2}{(\alpha_\theta-1)[{\alpha_\theta}\cot\frac{\xi-\mu}{2}+Q(z)]}\left(\frac{A+i}{A-z}\frac{I-\mathcal{C}_{\theta,\omega}}{2}\cdot,g_i^-(\theta)\right)\frac{A-i}{A-z}g_i^-(\theta).
\end{eqnarray*}

It follows from (\ref{ss1}), (\ref{e4}), (\ref{e13}), (\ref{new14}),
and (\ref{dede1}) that
$$
(I+\mathcal{C}_{\theta,\omega}^*)g_i^+(\theta)=(I+\mathcal{C}_{1/\theta,\omega})g_i^+(\theta)=2\alpha_\theta\left(e_{++}-\frac{1-\alpha_\theta}{\beta_\theta}e^{i\omega}e_{+-}\right)=2\alpha_\theta{g_i^+(1/\theta)}.
$$
Therefore, for any $f\in\mathfrak{H}$,
$$
\left(\frac{A+i}{A-z}\frac{I+\mathcal{C}_{\theta,\omega}}{2}f,g_i^+(\theta)\right)=\left(\frac{I+\mathcal{C}_{\theta,\omega}}{2}\frac{A+i}{A-z}f,g_i^+(\theta)\right)=\alpha_\theta\left(\frac{A+i}{A-z}f,g_i^+(1/\theta)\right).
$$

Similarly,
$(I-\mathcal{C}_{\theta,\omega}^*)g_i^-(\theta)=2\alpha_{\theta}g_i^-(1/\theta)$
and
$$
\left(\frac{A+i}{A-z}\frac{I-\mathcal{C}_{\theta,\omega}}{2}f,g_i^-(\theta)\right)=\alpha_\theta\left(\frac{A+i}{A-z}f,g_i^-(1/\theta)\right).
$$
Substituting the obtained expressions into the above resolvent
formula and taking the evident relation $
\frac{\alpha_\theta\beta_\theta^2}{\alpha_\theta-1}=\frac{\alpha_\theta(\alpha_\theta+1)\beta_\theta^2}{\alpha_\theta^2-1}
=\alpha_\theta(\alpha_\theta+1)$ into account, we complete the proof
of Theorem \ref{te14}. \rule{2mm}{2mm}
\begin{ccc}\label{new3}
Let the spectrum of $A\in\Upsilon$ be purely essential (i.e.,
$\sigma(A)=\sigma_{\mathrm{ess}}(A)$) and let $A_{M(U)}$ be an
arbitrary $J$-self-adjoint extension of $A_{\mathrm{sym}}$ with
${\mathcal{C}}_{\theta,\omega}$-symmetry. Then the essential
spectrum of $A_{M(U)}$ coincides with $\sigma_{\mathrm{ess}}(A)$ and
the discrete spectrum $\sigma_{\mathrm{disc}}(A_{M(U)})$ is
determined as the solutions of the equation
\begin{equation}\label{new4}\fl
\left[{\alpha_\theta}\tan\frac{\xi+\mu}{2}-Q(z)\right]\cdot\left[{\alpha_\theta}\cot\frac{\xi-\mu}{2}+Q(z)\right]=0,
\quad {z}\in{\mathbb{R}}\setminus\sigma_{\mathrm{ess}}(A),
\end{equation}
where
$Q(z)=2\left(\frac{1+zA}{A-z}e_{++},({\alpha_\theta}I-{\beta_\theta}R_\omega)e_{++}\right)$.
\end{ccc}

The proof of Corollary \ref{new3} immediately follows from the
resolvent formula in Theorem \ref{te14} if one takes into account
the following arguments: 1. $A$ and $A_{M(U)}$ are self-adjoint in
$\mathfrak{H}$ with respect to inner product
$(\cdot,\cdot)_\mathcal{C}$ (Subsection 2.2) and they are reduced by
the decomposition ${\mathfrak{H}}={\mathfrak
L}_+^{\theta,\omega}\oplus_{\mathcal C}{\mathfrak L}_-^{\theta,
\omega}$ (see (\ref{dede4})); 2. The second and the third parts on
the right-hand side of the resolvent formula belong to ${\mathfrak
L}_+^{\theta,\omega}$ and ${\mathfrak L}_-^{\theta,\omega}$,
respectively (since $g_i^{\pm}(\theta)\in{\mathfrak
L}_\pm^{\theta,\omega}$).

\setcounter{equation}{0}
\section{Examples}
\subsection{Schr\"{o}dinger operator with general
 zero-range potential.}
A one-dimensional Schr\"{o}dinger operator corresponding to a
general zero-range potential at the point $x=0$ can be given by the
expression
\begin{equation}\label{e23}\fl
-\frac{d^2}{dx^2}+
t_{11}<\delta,\cdot>\delta+t_{12}<\delta',\cdot>\delta+
 t_{21}<\delta,\cdot>\delta'+t_{22}<\delta',\cdot>\delta',
\end{equation}
where $\delta$ and $\delta'$ are, respectively, the Dirac
$\delta$-function and its derivative (with support at $0$) and
$t_{ij}$ are complex numbers.

The standard approach \cite{AL1} enables one to consider an operator
realization ${A}_T$ ($T=\|t_{ij}\|$) of (\ref{e23}) in
$L_2(\mathbb{R})$ by setting
\begin{equation}\label{lesia40}\fl
{A}_T={A}_{\mathrm{reg}}\upharpoonright{\mathcal{D}({A}_T)}, \
 \mathcal{D}({A}_T)=\{\,f\in{W_2^2}(\mathbb{R}\backslash\{0\}) :
 {A}_{\mathrm{reg}}f\in{L_2(\mathbb{R})}\},
\end{equation}
where the regularization of (\ref{e23}) onto
${W_2^2}(\mathbb{R}\backslash\{0\})$ has the form
$$
{A}_{\mathrm{reg}}=-\mathrm{\frac{d^2}{dx^2}}+t_{11}<\delta_{ex},\cdot>\delta+t_{12}<\delta_{ex}',\cdot>\delta+
t_{21}<\delta_{ex},\cdot>\delta'+t_{22}<\delta_{ex}',\cdot>\delta'.
$$
Here $-\mathrm{\frac{d^2}{dx^2}}$ acts on
${W_2^2}(\mathbb{R}\backslash\{0\})$ in the distributional sense and
$$
 <\delta_{ex}, f>=\frac{f(+0)+f(-0)}{2}, \quad <\delta_{ex}', f>=-\frac{f'(+0)+f'(-0)}{2}
$$
for all $f\in{{W_2^2}(\mathbb{R}\backslash\{0\})}$.

An operator realization $A_T$ of (\ref{e23}) is an intermediate
extension (i.e.,
$A_{\mathrm{sym}}\subset{A_T}{\subset}A_{\mathrm{sym}}^*$) of the
symmetric operator
\begin{equation}\label{as17}
A_{\mathrm{sym}}=-\frac{d^2}{dx^2}\upharpoonright{\{u\in{W}_2^2(\mathbb{R})
:\, u(0)=u'(0)=0\}}
\end{equation}
associated with (\ref{e23}).

Let ${\mathcal{P}}$ be the space parity operator
(${\mathcal{P}}f(x)=f(-x)$) in $L_2(\mathbb{R})$. The family of
${\mathcal{P}}$-self-adjoint operator realizations $A_T$ of
(\ref{e23}) is distinguished by the conditions $ t_{11},
t_{22}\in\RR, \  t_{21}=-\overline{t_{12}} $ imposed on the entries
$t_{ij}$ of $T$ \cite{AK}. Another description of
${\mathcal{P}}$-self-adjoint extensions of $A_{\mathrm{sym}}$ can be
found in \cite{A3}.

Let us consider the fundamental symmetry $Rf(x)=\sign (x)f(x)$ in
$L_2(\mathbb{R})$. Obviously, ${\mathcal{P}}{R}=-R{\mathcal{P}}$.
Since the operator $A_{\mathrm{sym}}$ in (\ref{as17}) has the
deficiency indices $<2,2>$ and commutes with $J\equiv{\mathcal{P}}$
and $R$ one can define the family of
${\mathcal{C}}_{\theta,\omega}$-symmetries by (\ref{e34}) and
(\ref{e13}).
\begin{ttt}\label{t267}
A ${\mathcal{P}}$-self-adjoint operator realization $A_T$ of
(\ref{e23}) has the property of ${\mathcal C}$-symmetry, where
${\mathcal C}$ commutes with $A_{\mathrm{sym}}$ if and only if there
exist $\theta>0, \omega, \phi, \xi\in[0,2\pi)$ such that the matrix
$T$ has the form
$$
T=\frac{2}{\Delta}\left(\begin{array}{cc}
\sqrt{2}(\alpha_\theta\sin\phi-\sqrt{1+\beta_\theta^2\sin^2\phi}\cos\xi)
& -\beta_\theta\cos\phi{e^{-i\omega}} \\
\beta_\theta\cos\phi{e^{i\omega}} &
-\sqrt{2}(\alpha_\theta\sin\phi-\sqrt{1+\beta_\theta^2\sin^2\phi}\sin\xi)
\end{array}\right),
$$
where
$\Delta=\alpha_\theta(\cos\phi-\sin\phi)+\sqrt{1+\beta_\theta^2\sin^2\phi}(\cos\xi+\sin{\xi})$.
In that case $A_T$  has ${\mathcal{C}}_{\theta,\omega}$-symmetry.
\end{ttt}

\emph{Proof.} Since $A_{\mathrm{sym}}$ is nonnegative, the existence
of a ${\mathcal C}$-symmetry for $A_T$, where ${\mathcal
C}A_{\mathrm{sym}}=A_{\mathrm{sym}}{\mathcal C}$ is equivalent to
the ${\mathcal{C}}_{\theta,\omega}$-symmetry of $A_T$ for some
choice of $\theta>0$ and $\omega\in[0,2\pi)$ (see Theorem
\ref{p14}).

The family of ${\mathcal{P}}$-self-adjoint extensions $A_{M(U)}$ of
$A_{\mathrm{sym}}$ having the property of
${\mathcal{C}}_{\theta,\omega}$-symmetry is described in Theorem
\ref{k18}. Therefore, the proof of Theorem \ref{t267} consists in
finding direct connections between the parameters of matrices $U$ in
(\ref{as34}) and the entries $t_{ij}$ of $T$ providing the equality
$A_T=A_{M(U)}$. To do this we note that the defect subspaces
$\mathfrak{N}_{+i}$ and $\mathfrak{N}_{-i}$ of $A_{\mathrm{sym}}$
coincide, respectively, with the linear spans of functions $<h_{1+},
h_{2+}>$ and $<h_{1-}, h_{2-}>$, where
\begin{equation}\label{aaa7}\fl
h_{1\pm}(x)=\left\{\begin{array}{cc}
 e^{i\tau_\pm{x}}, & x>0  \\
 e^{-i\tau_\pm{x}}, & x<0
 \end{array}\right.    \hspace{10mm}
 h_{2\pm}(x)=\left\{\begin{array}{cc}
 -e^{i{\tau_{\pm}}x}, & x>0  \\
 e^{-i\tau_\pm{x}}, & x<0
 \end{array}\right.
\end{equation}
and $\tau_\pm=\pm\frac{1}{\sqrt{2}}+i\frac{1}{\sqrt{2}}$ \
($\tau_\pm^2=\pm{i}$).

Since ${\mathcal P}h_{1\pm}=h_{1\pm}$ and ${\mathcal
P}h_{2\pm}=-h_{2\pm}$, the orthonormal basis $\{e_{\pm\pm}\}$ of the
Hilbert space
$\mathfrak{M}=\mathfrak{N}_{-i}\dot{+}\mathfrak{N}_{i}$ (see
(\ref{aaa2})) takes the form:
\begin{equation}\label{aaa4}
e_{++}=\alpha{h_{1+}}, \quad e_{+-}=\alpha{h_{2+}}, \quad
e_{-+}=\alpha{h_{1-}}, \quad e_{--}=\alpha{h_{2-}},
\end{equation}
where $\alpha=2^{-3/4}$ is a normalizing constant.

Let a ${\mathcal{P}}$-self-adjoint operator $A_T$ be determined by
(\ref{lesia40}). It is known \cite{AK} that  $A_T$ can be described
as the restriction of $A_{\mathrm{sym}}^*$ onto
\begin{equation}\label{e22}
\mathcal{D}(A_T)=\{f\in{W_2^2}(\mathbb{R}\backslash\{0\}) :\,
T\Gamma_0{f}=\Gamma_1{f}\ \},
\end{equation}
where $ \Gamma_0f=\displaystyle{\frac{1}{2}}\left(\begin{array}{c}
 f(+0)+f(-0) \\
-f'(+0)-f'(-0)
\end{array}\right)$ \quad and \quad  $\Gamma_1f=\left(\begin{array}{c}
f'(+0)-f'(-0) \\
f(+0)-f(-0)
\end{array}\right).
$

It follows from (\ref{e6}) and Theorem \ref{k18} that $A_T$ has
${\mathcal{C}}_{\theta,\omega}$-symmetry if and only if ${\mathcal
D}(A_T)={\mathcal D}(A_{M(U)})={\mathcal
D}(A_{\mathrm{sym}})\dot{+}M(U)$, where $M(U)$ is the linear span of
\begin{equation}\label{aaa6}
\begin{array}{c}
d_{1}=e_{++}-\frac{\beta_\theta}{\alpha_\theta}\cos\phi{e^{i(\phi+\omega)}}e_{+-}+\frac{1}{\alpha_\theta}\sqrt{1+\beta_\theta^2\sin^2\phi}{e^{i(\phi+\xi)}}e_{-+},
\\
d_{2}=e_{--}-\frac{1}{\alpha_\theta}\sqrt{1+\beta_\theta^2\sin^2\phi}{e^{i(\phi-\xi)}}e_{+-}-\frac{\beta_\theta}{\alpha_\theta}\cos\phi{e^{i(\phi-\omega)}}e_{-+}
\end{array}
\end{equation}

The boundary values $\Gamma_i{d_1}$ and $\Gamma_i{d_2}$ ($i=0,1$)
can easily be calculated with the help of (\ref{aaa4}). Substituting
these values into (\ref{e22}) instead of $\Gamma_if$ one derives a
system of linear equations with respect to $t_{ij}$. Its solution
(the matrix $T$ in Theorem \ref{t267}) gives the general form of all
$T$ such that $A_T=A_{M(U)}$. Only in this case the operator $A_T$
has ${\mathcal{C}}_{\theta,\omega}$-symmetry. Theorem \ref{t267} is
proved. \rule{2mm}{2mm}

Combining the description of $\Upsilon$ given in Proposition
\ref{k17} with formulas (\ref{aaa4}) and (\ref{aaa6}) leads to the
conclusion that a ${\mathcal{P}}$-self-adjoint extension $A_{M(U)}$
of $A_{\mathrm{sym}}$ belongs to $\Upsilon$ if and only if
$$
{\mathcal D}(A_{M(U)})=\{\,f\in{W_2^2}(\mathbb{R}\backslash\{0\}) :
 f(+0)=cf'(+0); \ f(-0)=-cf'(-0)\ \},
$$
where $c\in\mathbb{R}\cup\{\infty\}$. So, operators from $\Upsilon$
are characterized by separated boundary conditions and they are just
the second derivative self-adjoint operators on the half-lines. In
particular, the operator $A\in\Upsilon$ which has been used in
Theorem \ref{te14} corresponds to the case $c=0$, i.e.,
$$
{\mathcal D}(A)=\{\,f\in{W_2^2}(\mathbb{R}\backslash\{0\}) :
 f(+0)=0; \ f(-0)=0\ \}.
$$
This operator is the Friedrichs extension of $A_{\mathrm{sym}}$ and
the spectrum of $A$ is purely absolutely continuous and it coincides
with $[0,\infty)$. According to Corollary \ref{new3}, the discrete
spectrum of an arbitrary ${\mathcal{P}}$-self-adjoint extension
$A_{M(U)}$ is determined by (\ref{new4}), where $Q(z)$ can be
calculated in an explicit form with the use of (\ref{aaa7}) and
(\ref{aaa4}): $Q(z)=k(z)\alpha_\theta$, where
$$
k(z)=\frac{4\sqrt{2}}{\pi}\int_0^\infty\frac{y^2(1+zy^2)}{(y^2-z)(y^4+1)}dy.
$$
Therefore, $A_{M(U)}$ has a negative eigenvalue $z$ if and only if
\begin{equation}\label{aaa9}
\left[\tan\frac{\xi+\mu}{2}-k(z)\right]\cdot\left[\cot\frac{\xi-\mu}{2}+k(z)\right]=0,
\end{equation}
where $\mu=\mu(\theta, \phi)$ is determined by (\ref{as14}). The
formula (\ref{aaa9}) does not depend on $\omega$ in (\ref{as34}).
This means that the discrete spectrum of $A_{M(U)}$
$(U=U(\theta,\omega,\psi,\xi))$ does not depend on the choice of
$\omega$.

\subsection{One dimensional Dirac operator with point perturbation.}
Let us consider the free Dirac operator $D$ in the space
$L_2(\mathbb{R})\otimes{\mathbb{C}}^2$:
$$
D=-ic\frac{d}{dx}\otimes{\sigma_1}+\frac{c^2}{2}\otimes\sigma_3,
\quad \mathcal{D}(D)={W}_2^1(\mathbb{R})\otimes{\mathbb{C}}^2,
$$
where $\sigma_1$, \ $\sigma_3$ are Pauli matrices (see
(\ref{pauli})) and $c>0$ denotes the velocity of light.

The symmetric Dirac operator
$A_{\mathrm{sym}}=D\upharpoonright{\{u\in{W}_2^1(\mathbb{R})\otimes{\mathbb{C}}^2
:\, u(0)=0\}}$ has the deficiency indices $<2,2>$ \cite{AL,GS} and
it commutes with the fundamental symmetry
$J={\mathcal{P}}\otimes{\sigma_3}$ in
$L_2(\mathbb{R})\otimes{\mathbb{C}}^2$. Here
$u(\cdot)=\left(\begin{array}{c} u_1(\cdot) \\
u_2(\cdot)
\end{array} \right)\in{W}_2^1(\mathbb{R})\otimes{\mathbb{C}}^2$.

The defect subspaces $\mathfrak{N}_{i}$ and $\mathfrak{N}_{-i}$ of
$A_{\mathrm{sym}}$ coincide, respectively, with the linear spans of
the functions $<h_{1+}, h_{2+}>$ and $<h_{1-}, h_{2-}>$, where
\begin{equation}\label{aaa14}
h_{1\pm}(x)=\left(\begin{array}{c}
  -ie^{{\mp{i}t}} \\
 \sign(x)
 \end{array}\right){e^{i\tau|x|}}, \quad
 h_{2\pm}(x)=\sign(x)h_{1\pm}(x),\quad x\in\RR,
\end{equation}
$\tau=\frac{i}{c}\sqrt{\frac{c^4}{4}+1}$, and
$e^{it}:=\left(\frac{c^2}{2}-i\right)\left(\sqrt{\frac{c^4}{4}+1}\right)^{-1}$.

Since $Jh_{1\pm}=h_{1\pm}$ and $Jh_{2\pm}=-h_{2\pm}$, the
orthonormal basis $\{e_{\pm\pm}\}$ of the Hilbert space
$\mathfrak{M}=\mathfrak{N}_{-i}\dot{+}\mathfrak{N}_{i}$ (see
(\ref{aaa2})) takes the form:
\begin{equation}\label{aaa19}
e_{++}=\alpha{h_{1+}}, \quad e_{+-}=\alpha{h_{2+}}, \quad
e_{-+}=\alpha{h_{1-}}, \quad e_{--}=\alpha{h_{2-}}
\end{equation}
where $\alpha$ is a normalizing constant providing
$\|e_{\pm\pm}\|_{\mathfrak{M}}=1$.

The adjoint operator
$A_{\mathrm{sym}}^*=-i\frac{d}{dx}\otimes{\sigma_1}+{m}\otimes\sigma_3$
is defined on the domain
$\mathcal{D}(A_{\mathrm{sym}}^*)={W}_2^1(\mathbb{R}\setminus\{0\})\otimes{\mathbb{C}}^2$
and an arbitrary $J$-self-adjoint extension $A_{M(U)}$ of
$A_{\mathrm{sym}}$ is the restriction of $A_{\mathrm{sym}}^*$ onto
${\mathcal D}(A_{M(U)})={\mathcal{D}}(A_{\mathrm{sym}})\dot{+}M(U)$,
where $M(U)$ is defined by  (\ref{e6}) and (\ref{aaa5}) with
$e_{\pm\pm}$ determined by (\ref{aaa19}). Other descriptions of
$J$-self-adjoint extensions of $A_{\mathrm{sym}}$ can be found in
\cite{AL,BD,GS}.

To construct the family of
${\mathcal{C}}_{\theta,\omega}$-symmetries for $J$-self-adjoint
extensions $A_{M(U)}$ one needs to find a fundamental symmetry $R$
in $L_2(\mathbb{R})\otimes{\mathbb{C}}^2$ such that
$$
J{R}=-RJ \quad \mbox{and} \quad A_{\mathrm{sym}}R=RA_{\mathrm{sym}}.
$$
Obviously, these relations are satisfied for $R=\sign(x)I$. In that
case one can define the collection of
${\mathcal{C}}_{\theta,\omega}$-symmetries by (\ref{e13}). According
to Theorem \ref{k18}, a family of $J$-self-adjoint extensions
$\{A_{M(U)}\}$ having at least one
${\mathcal{C}}_{\theta,\omega}$-symmetry is described by subspaces
$M(U)=<d_1,d_2>$, where $d_i$ are determined by (\ref{aaa6}) and
(\ref{aaa19}). In the particular case $A_{M(U)}\in\Upsilon$ (i.e.,
$A_{M(U)}$ commutes with any ${\mathcal{C}}_{\theta,\omega}$),
relation (\ref{aaa10}) must be used instead of (\ref{aaa6}). A
routine calculation gives $A_{M(U)}\in\Upsilon$ if and only if
$$
{\mathcal
D}(A_{M(U)})=\left\{\,f\in{W}_2^1(\mathbb{R}\setminus\{0\})\otimes{\mathbb{C}}^2
:
\begin{array}{c}
{i}\cos(\frac{\xi}{2}+\frac{\pi}{4})f_1(+0)=\cos(t+\frac{\xi}{2}+\frac{\pi}{4})f_2(+0) \\
-{i}\cos(\frac{\xi}{2}+\frac{\pi}{4})f_1(-0)=\cos(t+\frac{\xi}{2}+\frac{\pi}{4})f_2(-0)
 \end{array} \right\},
$$
where $t$ is determined in (\ref{aaa14}) and $\xi\in[0,2\pi)$.
Hence, as in the case of a Schr\"{o}dinger operator, the elements of
$\Upsilon$ are characterized by separated boundary conditions.  The
operator $A\in\Upsilon$ in the resolvent formula (see Theorem
\ref{te14}) corresponds to the case $\xi=\frac{\pi}{2}$, i.e.,
$$
{\mathcal
D}(A)=\left\{\,f\in{W}_2^1(\mathbb{R}\setminus\{0\})\otimes{\mathbb{C}}^2
: \ f_2(+0)=f_2(-0)=0 \right\}
$$
($\cos(t+\frac{\pi}{2})\not=0$ by the definition of $t$). Since
$$
A^2=-c^2\frac{d^2}{dx^2}+\frac{c^4}{4}, \quad {\mathcal
D}(A^2)=\left\{\,f\in{W}_2^2(\mathbb{R}\setminus\{0\})\otimes{\mathbb{C}}^2
: \ \begin{array}{c} f_1'(+0)=f_1'(-0)=0 \\
f_2(+0)=f_2(-0)=0
\end{array}
\right\}$$ the spectrum of $A$ is purely absolutely continuous and
it coincides with $(-\infty, -c^2/2]\cup[c^2/2, \infty)$.

Let $A_{M(U)}$ be a $J$-self-adjoint extension of $A_{\mathrm{sym}}$
with ${\mathcal{C}}_{\theta,\omega}$-symmetry. Then $A_{M(U)}$ turns
out to be self-adjoint in $L_2(\mathbb{R})\otimes{\mathbb{C}}^2$
with respect to the inner product
$(\cdot,\cdot)_{{\mathcal{C}}_{\theta,\omega}}$. The corresponding
resolvent formula is given in Theorem \ref{te14}; the essential
spectrum of $A_{M(U)}$ coincides with $(-\infty, -c^2/2]\cup[c^2/2,
\infty)$ and its bound states $z\in(-c^2/2, c^2/2)$ can be found as
solutions of (\ref{new4}).

\section{Conclusions}
In the present paper von Neumann's self-adjoint extension technique
for symmetric operators has been reshaped to provide
\emph{$J$-self-adjoint extensions} of symmetric operators with
arbitrary but equal deficiency indices $<n,n>$, \ $n\in \NN \cup
\infty$. The crucial role is played by a bijection between the
resulting family of $J$-self-adjoint operators and hypermaximal
neutral subspaces of the defect Krein space. It is proven that the
$\cC$ operators of the resulting Hamiltonians leave the defect Krein
spaces invariant. For $J$-self-adjoint extensions of symmetric
operators with deficiency indices $<2,2>$ the parametrization of the
$\cC$-operator family is worked out in detail and Krein type
resolvent formulas are constructed. The technique is exemplified on
1D pseudo-Hermitian Schr\"odinger and Dirac Hamiltonians with
complex point-interaction potentials.

Due to their specific structure, Hamiltonians  obtained as
$J$-self-adjoint extensions of symmetric operators provide an
excellent playing ground for studies on the Krein-space related
features of pseudo-Hermitian and $\cP\cT-$symmetric operators. The
advantages of such model Hamiltonians have their origin in the
following properties. For sufficiently simple symmetric differential
operators the models remain exactly solvable. They have rich
parameter spaces which are bijectively related to the hypermaximal
neutral subspaces of the defect Krein spaces of the symmetric
operators. As differential operators the resulting pseudo-Hermitian
Hamiltonians possess, in general, much richer spectra than simple
matrix Hamiltonians, i.e. apart from discrete spectra they will have
continuous  and, possibly,  residual spectra. Corresponding
resolvent studies can be carried out in full detail with exact
results. In this way these Hamiltonians have the capability to
provide some deeper insights into the structural subtleties of
pseudo-Hermitian and $\cP\cT-$symmetric quantum theories.

\section{Acknowledgments}UG acknowledges support from DFG via the
Collaborative Research Center SFB 609. SK thanks the Saxon Ministry
of Sciences, grant 4-7531.50-04-844-08/7 and DFG 436 UKR 113/88/0-1
for support and Forschungszentrum Dresden-Rossendorf for warm
hospitality.

 \end{document}